\documentclass[english,manuscript,screen]{acmart}
\PassOptionsToPackage{ruled}{algorithm2e}
\usepackage[LGR,T1]{fontenc}
\usepackage[latin9]{inputenc}
\usepackage{array}
\usepackage{booktabs}
\usepackage{bbding}
\usepackage{multirow}
\usepackage{algorithm2e}
\usepackage{graphicx}
\usepackage{rotating}

\makeatletter

\acmDOI{XXXXXXX.XXXXXX}

\DeclareRobustCommand{\greektext}{%
  \fontencoding{LGR}\selectfont\def\encodingdefault{LGR}}
\DeclareRobustCommand{\textgreek}[1]{\leavevmode{\greektext #1}}
\ProvideTextCommand{\~}{LGR}[1]{\char126#1}

\newcommand{\lyxmathsym}[1]{\ifmmode\begingroup\def\b@ld{bold}
  \text{\ifx\math@version\b@ld\bfseries\fi#1}\endgroup\else#1\fi}

\providecommand{\tabularnewline}{\\}

 \setcopyright{none}

\SetAlFnt{\small}
\SetAlCapFnt{\small}
\SetAlCapNameFnt{\small}
\SetAlCapHSkip{0pt}
\IncMargin{-\parindent}

\makeatother

\usepackage{babel}
\begin{document}
\title[PERFORMANCE MODELING OF PUBLIC PERMISSIONLESS BLOCKCHAINS: A SURVEY]{PERFORMANCE MODELING OF PUBLIC PERMISSIONLESS BLOCKCHAINS: A SURVEY}
\author{Molud Esmaili}
\affiliation{\institution{University of South Florida}\country{USA}}
\author{Ken Christensen}
\affiliation{\institution{University of South Florida}\country{USA}}
\begin{abstract}
Public permissionless blockchains facilitate peer-to-peer digital
transactions, yet face performance challenges specifically minimizing
transaction confirmation time to decrease energy and time consumption
per transaction. Performance evaluation and prediction are crucial
in achieving this objective, with performance modeling as a key solution
despite the complexities involved in assessing these blockchains.
This survey examines prior research concerning the performance modeling
blockchain systems, specifically focusing on public permissionless
blockchains. Initially, it provides foundational knowledge about these
blockchains and the crucial performance parameters for their assessment.
Additionally, the study delves into research on the performance modeling
of public permissionless blockchains, predominantly considering these
systems as bulk service queues. It also examines prior studies on
workload and traffic modeling, characterization, and analysis within
these blockchain networks. By analyzing existing research, our survey
aims to provide insights and recommendations for researchers keen
on enhancing the performance of public permissionless blockchains
or devising novel mechanisms in this domain.
\end{abstract}

\end{CCSXML}

\maketitle

\section{Introduction}

A distributed ledger known as blockchain is formed by linking a sequence
of blocks within a peer-to-peer network \citep{33}. The foundational
concept of blockchain was initially introduced by Harber et al. in
1991 \citep{2}. Their paper outlines the idea of a distributed ledger,
where each timestamp contains a hash of the previous timestamp. In
2008, Nakamoto presented a paper on Bitcoin \citep{3}, which served
as the pioneering implementation of blockchain as a peer-to-peer electronic
cash system. Reaching an agreement within a blockchain network is
a critical undertaking. In this regard, numerous consensus mechanisms
have been designed and developed to maintain validity in a trustless
and unreliable network, and the variety of these mechanisms continues
to grow. The most renowned consensus approach is Proof of Work (PoW)
\citep{3}, initially introduced by Nakamoto and employed in Bitcoin.
However, due to concerns regarding throughput, energy efficiency,
and mining equipment cost limitations associated with PoW, alternative
consensus mechanisms like Proof of Stake (PoS) have been proposed.
PoS relies on the concept of selecting the next block creator through
a combination of random selection, the stake amount owned by a validator,
and stake's age or the duration it has been held by the validator,
enhancing scalability. This idea was first introduced in 2012 with
the Peercoin cryptocurrency \citep{4} and later utilized in other
projects such as Nxt \citep{5} and Blackcoin \citep{6}. An improvement
to PoS is Delegated Proof of Stake (DPoS), introduced by Larimer in
2014 \citep{7}, wherein nodes choose representatives through voting
to validate blocks. Another consensus mechanism that can be employed
in blockchain networks is Practical Byzantine Fault Tolerance (PBFT),
which was introduced by Castro in 1999 \citep{8} to address the Byzantine
generals' problem \citep{9}. Over time, several other variants of
these mechanisms have been proposed to enhance the consensus process
in blockchain networks. Examples of such variants include Proof of
Elapsed Time \citep{10}, Delegated Byzantine Fault Tolerance \citep{11},
Proof of Activity \citep{12}, and Proof of Burn \citep{13}. Blockchain
extends beyond cryptocurrency and finds applications in different
industries like healthcare \citep{17}, supply chain management \citep{149},
digital marketplaces \citep{146}, streaming platforms \citep{46}
and more \citep{26}.

Blockchain systems can be broadly categorized as either permissioned
or permissionless. Regarding a permissioned blockchain, only authorized
users are permitted to serve as consensus nodes and access the blockchain's
data. Conversely, in a permissionless blockchain, any user has the
freedom to join or exit the network \citep{14}. The other categorization
for blockchain systems is as public and private blockchains. Public
blockchains are openly accessible, permitting anyone to view the blockchain\textquoteright s
data. In contrast, private ledgers restrict access solely to those
who were approved beforehand \citep{15}. It's worth noting that existing
public permissionless blockchain systems, primarily based on variants
of PoW and PoS, experience performance issues such as limited throughput,
high communication overhead, and transaction confirmation latency
\citep{16}.

Cryptocurrencies, specifically Bitcoin, have much slower transaction
rates compared to VISA's network which handles thousands per second
\citep{160}. Addressing this latency is crucial for widespread cryptocurrency
adoption. To address the performance-related issues of blockchain
systems, numerous researchers have focused on the performance evaluation
of private and public blockchain systems. Furthermore, extensive research
has been conducted on scalability challenges and potential solutions
in public blockchains. Table 1 provides a classification of survey
papers that reviewed research on evaluating and improving performance
and scalability in blockchain systems.

The pioneering studies addressing the bulk service queuing problem,
namely \citep{52,53,139,140,141,142} drew inspiration from Kendall's
fundamental research in 1953 \citep{47}. Simulations and analytical
modeling are now commonly used approaches for understanding the features
of blockchain systems \citep{71}. Furthermore, analytical modeling
can be applied when a mathematical model offers a closed-form solution,
providing a simplified representation of system operation through
a set of equations \citep{36}. Consequently, analytical modeling
approaches are highly attractive for blockchain analysis and evaluation,
leading to numerous researchers reviewing models to analyze blockchain
systems \citep{18,19,20,21}. The various approaches employed to evaluate
the performance of blockchain systems are depicted in Fig 1.
\begin{figure}
\includegraphics[scale=0.55]{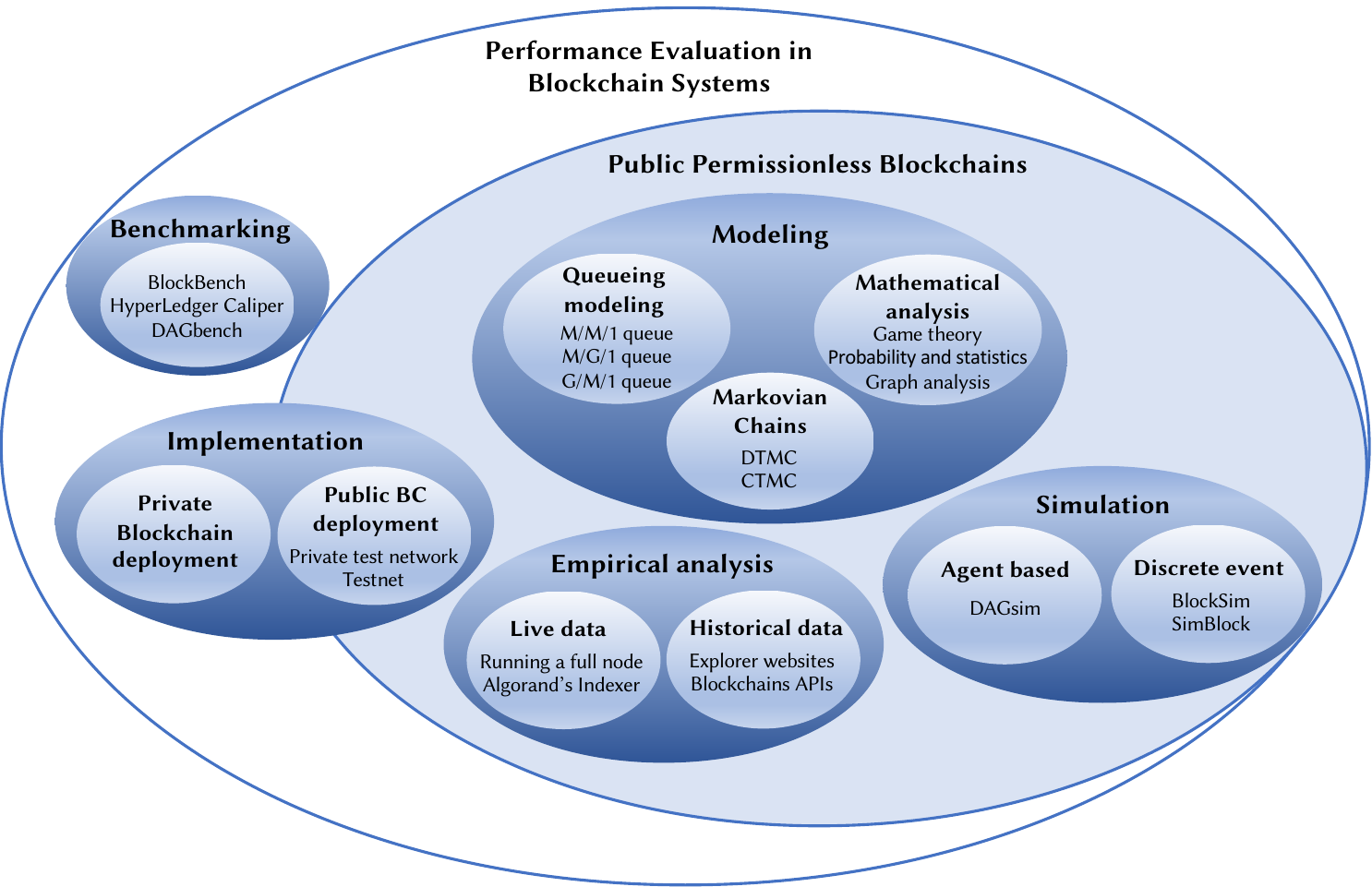}

\caption{Taxonomy of performance evaluation methods in blockchain systems}

\vspace{-.3cm} 
\end{figure}

\subsection{Motivation}

``Performance modeling'' involves developing abstract representations
or mathematical constructs that mimic how a system, application, or
process behaves in terms of performance parameters. These models help
predict and evaluate how well the systems will perform under various
workloads \citep{58}. When it comes to blockchain systems, performance
modeling approaches allow for the assessment and analysis of critical
performance metrics such as throughput and latency. Additionally,
given the substantial network traffic and decentralization in public
permissionless blockchains, the evaluation and enhancement of performance
metrics assume heightened significance. In addition, several new and
unreviewed blockchain performance models have been conducted in recent
years \citep{62,66,70,72,73,74,81,82}. 

Upon conducting an overview, as presented in Tables 1, it becomes
evident that a comprehensive survey of performance models of public
permissionless blockchain systems is still missing. Such a survey
would serve a dual purpose: (i) aiding researchers in selecting the
most suitable performance modeling approach to evaluate public permissionless
blockchains, and (ii) assisting researchers in characterizing and
analyzing the workload and traffic of public permissionless blockchains
for a more precise performance evaluation. This motivation has prompted
us to compile this survey, offering a recent review of this increasingly
important field. The related survey papers to our survey are summarized
in Table 1.
\begin{table}[h]
\caption{The related survey papers}
\vspace{-.2cm} 

\begin{tabular*}{1\linewidth}{@{\extracolsep{\fill}}lll>{\raggedright}p{1\textheight}}
\toprule 
What is surveyed & Ref & Year & Main contributions\tabularnewline
\midrule
\midrule 
{\footnotesize{}Consensus algorithms} & {\footnotesize{}\citep{22}} & {\small{}2017} & {\footnotesize{}Performance analysis of consensus mechanisms}\tabularnewline
{\footnotesize{}from a performance} & {\footnotesize{}\citep{23}} & {\small{}2018} & {\footnotesize{}Distributed ledgers, cryptography, consensus protocols,
and smart contracts}\tabularnewline
{\footnotesize{}perspective} & {\footnotesize{}\citep{24}} & {\small{}2019} & {\footnotesize{}Proof-of-Stake-based mechanisms}\tabularnewline
 & {\footnotesize{}\citep{25}} & {\small{}2019} & {\footnotesize{}Performance and security concerns of distributed ledgers,
cryptographies, consensus mechanisms}\tabularnewline
 & {\footnotesize{}\citep{170}} & {\small{}2020} & {\footnotesize{}Performance comparison of consensus mechanisms}\tabularnewline
 & {\footnotesize{}\citep{27}} & {\small{}2020} & {\footnotesize{}Performance and scalability comparative analysis}\tabularnewline
 & {\footnotesize{}\citep{28}} & {\small{}2020} & {\footnotesize{}Performance analysis of PoW, PoS, Pure PoS}\tabularnewline
 & {\footnotesize{}\citep{29}} & {\small{}2021} & {\footnotesize{}Performance and security analysis}\tabularnewline
 & {\footnotesize{}\citep{30}} & {\small{}2021} & {\footnotesize{}Performance analysis of PoW, PoS, DPoS, and PoA}\tabularnewline
 & {\footnotesize{}\citep{31}} & {\small{}2021} & {\footnotesize{}Performance analysis of recently proposed mechanisms}\tabularnewline
 & {\footnotesize{}\citep{32}} & {\small{}2022} & {\footnotesize{}Proof of Stake-based consensus algorithms}\tabularnewline
 & {\footnotesize{}\citep{33}} & {\small{}2023} & {\footnotesize{}Review and performance analysis of consensus mechanisms}\tabularnewline
\midrule
{\footnotesize{}Blockchain systems\textquoteright{}} & {\footnotesize{}\citep{34}} & {\small{}2019} & {\footnotesize{}Blockchain benchmarking tools and optimization methods}\tabularnewline
{\footnotesize{}performance and} & {\footnotesize{}\citep{35}} & {\small{}2020} & {\footnotesize{}Empirical performance evaluations of permissioned
blockchains}\tabularnewline
{\footnotesize{}scalability approaches} & {\footnotesize{}\citep{36}} & {\small{}2020} & {\footnotesize{}Performance evaluation approaches in blockchain}\tabularnewline
{\footnotesize{}and solutions} & {\footnotesize{}\citep{37}} & {\small{}2020} & {\footnotesize{}Consensus mechanisms performance evaluation criteria}\tabularnewline
 & {\footnotesize{}\citep{15}} & {\small{}2020} & {\footnotesize{}Performance evaluation approaches for blockchain systems}\tabularnewline
 & {\footnotesize{}\citep{39}} & {\small{}2018} & {\footnotesize{}Scalability solutions in blockchains}\tabularnewline
 & {\footnotesize{}\citep{40}} & {\small{}2019} & {\footnotesize{}Scalability solutions in blockchains}\tabularnewline
 & {\footnotesize{}\citep{41}} & {\small{}2020} & {\footnotesize{}Scalability solutions in blockchains}\tabularnewline
 & {\footnotesize{}\citep{42}} & {\small{}2020} & {\footnotesize{}Scalability solutions in blockchains}\tabularnewline
 & {\footnotesize{}\citep{43}} & {\small{}2020} & {\footnotesize{}Scalability solutions in blockchains}\tabularnewline
 & {\footnotesize{}\citep{44}} & {\small{}2021} & {\footnotesize{}Scalability solutions in blockchains}\tabularnewline
 & {\footnotesize{}\citep{45}} & {\small{}2021} & {\footnotesize{}Scalability and performance issues and solutions in
blockchains}\tabularnewline
\midrule
{\footnotesize{}Models of blockchains\textquoteright{}} & {\footnotesize{}\citep{18}} & {\small{}2020} & {\footnotesize{}Stochastic modeling approaches to analyze blockchain
systems}\tabularnewline
{\footnotesize{}features and phases} & {\footnotesize{}\citep{20}} & {\small{}2021} & {\footnotesize{}Modeling, theories, and tools to analyze blockchain
systems}\tabularnewline
 & {\footnotesize{}\citep{19}} & {\small{}2023} & {\footnotesize{}Models used for characterization of blockchain systems'
features}\tabularnewline
 & {\footnotesize{}\citep{21}} & {\small{}2023} & {\footnotesize{}Stochastic process, game theory, and ML methods used
to analyze blockchains}\tabularnewline
\bottomrule
\end{tabular*}\vspace{-.35cm} 
\end{table}

\subsection{Our Contribution}
\begin{itemize}
\item Our review encompasses an introduction to an exploration of selected
public permissionless consensus algorithms and an examination of performance
evaluation methods and criteria in blockchain systems.
\item We review and compare existing studies that employ modeling approaches
to investigate public permissionless blockchains' performance. We
analyze their advantages and limitations to offer guidance for future
researchers. 
\item Ultimately, we extensively explored research focused on proposing
models for blockchain network topologies, examining traffic and workload
characteristics, and initiating the development of benchmarking workloads.
\end{itemize}

\subsection{Organization}

This survey is structured as follows: In Section 2, we provide an
introduction to the fundamentals of selected public permissionless
blockchains, the performance evaluation of blockchains, and the criteria
for evaluating the performance of blockchains. Section 3 involves
a review of research focusing on the modeling of public permissionless
blockchain systems' performance. Moving to Section 4, we delve into
the examination of the studies conducted on workload modeling for
public permissionless blockchains. Section 5 outlines several challenges
and suggests potential future research directions, and lastly, Section
6 serves as the conclusion of this article.

\section{Background}

\subsection{Public permissionless blockchains}

As previously noted, existing public permissionless blockchains utilizing
PoW and PoS encounter scalability and performance challenges. We have
selected three well-known public permissionless blockchains, Bitcoin
\citep{3} as a PoW-based blockchain, Ethereum 2.0 \citep{167}based
on PoS, and Algorand \citep{144}as a combination of PoS and PBFT,
offering a condensed overview and a comparison of their performance
metrics.

\subsubsection{Proof of Work}

\paragraph{Bitcoin}

Consensus mechanism: Proof of Work (PoW) is Bitcoin's pioneering consensus
mechanism. Nodes compete to calculate block header hash functions
by finding a value that must be equal to or less than a given \textquotedbl nonce\textquotedbl{}
value. Validated blocks are broadcasted and appended to all the nodes'
chains. The time-consuming authentication process incentivizes miners
with mining rewards, consisting of the block reward and the included
transaction fees \citep{160}.

Bitcoin's data organization centers on transactions, blocks, and nodes.
Transactions encompass various fields like version number, flags,
inputs, outputs, and lock time. Blocks, each 1MB, contain a magic
number, block size, header, transaction counter, and transactions\citep{3}.
Each node in the Bitcoin network maintains a memory pool, known as
the Mempool, which serves as a temporary storage area for unconfirmed
transactions. These transactions have been broadcast to the network
but have not yet been included in a block by miners \citep{164}.
Bitcoin utilizes the Unspent Transaction Output (UTXO) set to monitor
unspent transaction outputs, which serve as inputs for new transactions.
Full Bitcoin nodes maintain a UTXO copy to validate and create transactions
efficiently, avoiding the need to review the entire blockchain \citep{165}.
Segregated Witness (SegWit) \citep{161} is a protocol upgrade implemented
in the Bitcoin network to improve the scalability and security of
Bitcoin transactions by separating the transaction signatures (witness
data) from the transaction data \citep{162}. SegWit increases block
capacity without altering the base block size but introduces a new
unit of \textquotedblleft Weight unit\textquotedblright{} and switches
the block limit size form 1 MB to 4 million weight units.

Bitcoin faces scalability issues due to block size limitations (1MB)
and block creation time (10 minutes), restricting transaction throughput.
This hinders the network's capacity to handle increased transaction
volume, posing a scalability challenge \citep{163}. The performance
parameters of Bitcoin are summarized and compared with those of several
other public permissionless blockchains in Table 2.

\subsubsection{Proof of Stake}

\paragraph{Ethereum 2.0.}

The Ethereum blockchain, originally relying on a proof-of-work mechanism
similar to Bitcoin, underwent a significant upgrade with Ethereum
2.0 released in 2022. This update aimed to enhance security, scalability,
and energy efficiency within the network \citep{166}. 

In Ethereum 2.0's proof-of-stake (PoS) consensus mechanism, time is
segmented into slots (each lasting 12 seconds) and epochs (comprising
32 slots). Validators are chosen at random to propose blocks in slots,
while a validator committee validates the proposed blocks' authenticity.
When a node initiates a transaction using its private key, it attaches
a gas fee as an incentive for validators to verify the sender's ETH
balance and signature. Once validated, the transaction enters the
node's local Mempool and gets broadcast to other nodes. Subsequently,
a randomly selected block proposer creates the next block, including
transactions from the Mempool \citep{167}.

In Ethereum 2.0's PoS, consensus clients utilize the LMD-GHOST algorithm
\citep{168} to resolve discrepancies among validators' perspectives.
LMD-GHOST identifies the fork with the most substantial historical
weight of attestations, ensuring agreement on the chain's head. Ethereum
also introduced the concept of accounts within its protocol, serving
as both the originator and recipient of transactions. Consequently,
transactions directly alter account balances rather than maintaining
a state, as seen in Bitcoin's UTXOs. This approach enables the transfer
of values, messages, and data among accounts, leading to state transitions
\citep{163}. 

Ethereum 2.0's performance limitation involves potential bottlenecks
in shard communication and data transfer, impacting overall scalability
despite its sharding improvements \citep{167}. Table 2 provides a
summary of the performance metrics for Ethereum 2.0.

\paragraph{Algorand.}

Algorand \citep{144} operates as a cryptocurrency utilizing a Byzantine
Agreement (BA) protocol, merged with Verifiable Random Functions \citep{169}.
This amalgamation selects users as committee members via a cryptographic
process to engage in the Byzantine Agreement for transaction consensus.
In contrast to Proof of Work, Algorand demands minimal computational
power and maintains a low probability of forking \citep{170}. Additionally,
once a block is integrated into the blockchain, it attains finality.

Algorand's consensus mechanism comprises several key steps: a) Block
Proposal: Each round allows anyone to propose a block. b) Random Participant
Selection: A small, stake-based group is chosen privately, ensuring
energy efficiency and decentralized involvement. c) Block Proposer
Election: From this group, one participant becomes the block proposer
for that round. d) Block Proposal and Voting: The proposer creates
a block, shared across the network, followed by nodes voting on its
acceptance using a secure voting scheme. e) Fast and Secure Consensus:
Algorand employs \textquotedbl BA{*} Byzantine Agreement\textquotedbl{}
for swift and secure agreement within a fixed number of rounds.

If the network agrees upon the proposed block, it becomes validated
as the subsequent block in the blockchain. Should the network reject
it, the procedure restarts in the subsequent round, featuring a fresh
random selection of participants and a different individual responsible
for proposing the block \citep{144}. 

Algorand faces performance limitations due to block propagation delay
and network latency. While its consensus mechanism ensures scalability
and security, the speed of block dissemination across the network
and synchronization can impact transaction throughput, potentially
causing delays in confirming transactions \citep{144}. Table 2 outlines
and compares the performance metrics of Algorand with several other
public permissionless blockchains.

\subsection{Performance evaluation criteria in blockchain systems}

In this subsection, we examined and compared performance evaluation
criteria and notable latencies across various popular public permissionless
blockchains; Bitcoin, Algorand, and Ethereum 2.0 illustrated in Table
2.

\begin{table}[h]
\vspace{-.1cm} 

\caption{performance criteria in the selected blockchain systems}
\vspace{-.2cm} 

\begin{tabular}{llllll}
\toprule 
\multicolumn{2}{l}{{\footnotesize{}Parameter}} & {\footnotesize{}Definition} & {\footnotesize{}BTC} & {\footnotesize{}ETH2.0} & {\footnotesize{}Algorand}\tabularnewline
\midrule
\midrule 
\multicolumn{2}{l}{{\footnotesize{}TPS}} & {\footnotesize{}The number of transactions processed per second} & {\footnotesize{}\ensuremath{\approx} 7} & {\footnotesize{}\ensuremath{\approx} 100,000} & {\footnotesize{}\ensuremath{\approx} 10,000}\tabularnewline
\midrule 
\multicolumn{2}{l}{{\footnotesize{}Transaction Confirmation Time }} & {\footnotesize{}The time interval between transaction creation and
transaction } & {\footnotesize{}\ensuremath{\approx} 10-30 min} & {\footnotesize{}\ensuremath{\approx} 5 min} & {\footnotesize{}\ensuremath{\approx} 4 sec}\tabularnewline
\multicolumn{2}{l}{} & {\footnotesize{}permanently addition to the blockchain} &  &  & \tabularnewline
\midrule 
\multicolumn{2}{l}{{\footnotesize{}Block Generation Time}} & {\footnotesize{}The time interval between creation times of two consecutive
blocks} & {\footnotesize{}\ensuremath{\approx} 10 min} & {\footnotesize{}\ensuremath{\approx} 12 sec} & {\footnotesize{}\ensuremath{\approx} 3.3 sec}\tabularnewline
\midrule 
\multicolumn{2}{l}{{\footnotesize{}Block Size }} & {\footnotesize{}The maximum amount of data can be included in a single
block} & {\footnotesize{}\ensuremath{\approx} 5000 tx} & {\footnotesize{}\ensuremath{\approx} 1500 tx} & {\footnotesize{}\ensuremath{\approx} 25000 tx}\tabularnewline
\midrule 
\multirow{10}{*}{\begin{turn}{90}
{\footnotesize{}Latencies}
\end{turn}} & {\footnotesize{}Consensus latency} & {\footnotesize{}The time it takes for the nodes to reach an agreement
on the validity} & {\footnotesize{}high} & {\footnotesize{}low} & {\footnotesize{}very low}\tabularnewline
 &  & {\footnotesize{}of transactions} &  &  & \tabularnewline
\cmidrule{2-6} \cmidrule{3-6} \cmidrule{4-6} \cmidrule{5-6} \cmidrule{6-6} 
 & {\footnotesize{}Finality latency} & {\footnotesize{}The time it takes for a transaction to be irreversibly
confirmed and } & {\footnotesize{}high} & {\footnotesize{}moderate} & {\footnotesize{}zero}\tabularnewline
 &  & {\footnotesize{}considered as ``Fsinalized''} &  &  & \tabularnewline
\cmidrule{2-6} \cmidrule{3-6} \cmidrule{4-6} \cmidrule{5-6} \cmidrule{6-6} 
 & {\footnotesize{}Waiting latency} & {\footnotesize{}The time a tx spends in a queue or pending state before
being } & {\footnotesize{}high} & {\footnotesize{}low} & {\footnotesize{}very low}\tabularnewline
 &  & {\footnotesize{}processed, validated, and added to the blockchain} &  &  & \tabularnewline
\cmidrule{2-6} \cmidrule{3-6} \cmidrule{4-6} \cmidrule{5-6} \cmidrule{6-6} 
 & {\footnotesize{}Tx/Block Propagation} & {\footnotesize{}The time it takes for newly created blocks or arrived
transactions to } & {\footnotesize{}high} & {\footnotesize{}low} & {\footnotesize{}moderate}\tabularnewline
 & {\footnotesize{}latency} & {\footnotesize{}be disseminated across the entire network} &  &  & \tabularnewline
\cmidrule{2-6} \cmidrule{3-6} \cmidrule{4-6} \cmidrule{5-6} \cmidrule{6-6} 
 & {\footnotesize{}Processing overheads} & {\footnotesize{}Computational power is needed for consensus mechanisms,
and} & {\footnotesize{}very high} & {\footnotesize{}low} & {\footnotesize{}low}\tabularnewline
 &  & {\footnotesize{}handling any other protocols in the system} &  &  & \tabularnewline
\bottomrule
\end{tabular}

\vspace{-.4cm} 
\end{table}

\section{Performance modeling of public permissionless blockchains}

In this section, our objective is to classify and review the research
conducted in the field of performance modeling of public permissionless
blockchains. Several stochastic models have been employed to assess
the performance and security vulnerabilities of blockchains. These
models encompass Markov chains, Queueing models, Stochastic Petri
nets, and other models such as Random graph model and stochastic network
model\citep{15}. Markov process has been utilized to model the general
features of blockchains, such as consensus algorithms and platforms
\citep{48,49,50,51}. Furthermore, stochastic models have been applied
to evaluate blockchain security in the work presented by \citep{54}.
When it comes to performance evaluation, Markov chains have been employed
to model the performance of private or permissioned blockchains like
PBFT \citep{57}, Raft \citep{48}, the tangle \citep{59}, and HLF
\citep{55,56}. Similarly, Stochastic Petri nets have found utility
in modeling the performance of private or permissioned blockchains,
as seen in the work on PBFT \citep{55,60}.

For the performance modeling of public permissionless blockchains,
queueing theory stands out as the most suitable modeling approach.
Because one of the most critical aspects of performance evaluation
in public blockchains is transaction latency and transaction confirmation
rate, which can be effectively modeled using queueing models. Therefore,
in this paper, we will concentrate on this specific modeling approach. 

\subsection{Performance modeling using queueing theory}

In the past decade, queueing theory has seen broad application in
modeling computer networks and also blockchain systems \citep{15}.
Within the context of a blockchain system, user-initiated transactions
must go through queues where miners or validators process them, add
them to a new block, validate them, and finally confirm them. In blockchain
networks, as you can see in Figure 2, various phases of consensus
and block generation can be represented as queueing systems, including
queues for arrival transactions, transaction confirmations, and block
verifications. Within queue theory, some important performance parameters,
including the expected number of transactions within the system, transaction
throughput, and average block generation time can be assessed. 

Multiple queue types have been examined and categorized using the
standardized Kendall's notation \citep{47}. In Kendall's notation,
queueing models are denoted as A/S/C, where A represents the inter-arrival
time for queue entries (input), S indicates the distribution of service
times (queue-discipline), and c represents the number of service channels
at the node (service mechanism). Common queue types utilized in blockchain
modeling include M/G/1, M/M/1, G/M/1, and GI/GI/1. In this notation,
M denotes Markovian behavior with Poisson arrival times, signifying
exponential inter-arrival times. G stands for general distribution,
GI denotes general independent distribution, and the last number represents
the count of servers in the system \citep{47}.
\begin{figure}[h]
\includegraphics[scale=0.45]{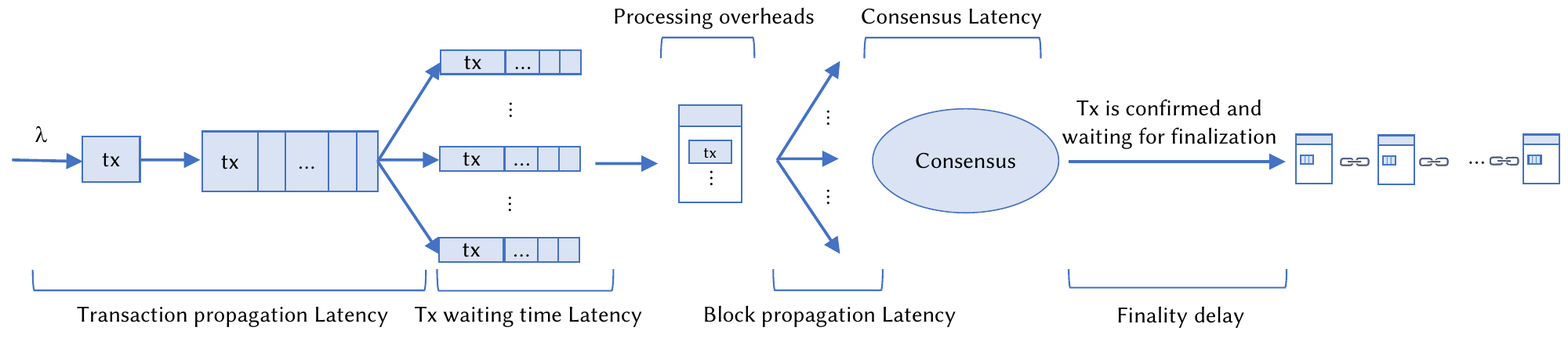}\vspace{-.3cm} \caption{The transaction confirmation process in blockchain systems}

\vspace{-.20cm} 
\end{figure}

In the following sections, we will introduce common types of queues
that are used in blockchain modeling and review the research that
has been conducted using each type of queue system.

\subsubsection{Modeling used M/G/1 queue.}

In 2016, Kasahara and Kawase \citep{65} developed a model representing
Bitcoin's transaction-confirmation process, utilizing a single-server
$M/G/1$ queue with batch service, a priority mechanism, and a non-work-conserving
discipline. The paper defines \textquotedbl Sojourn Time\textquotedbl{}
or ``Transaction-confirmation Time'' ($E[T]$) as the period between
a user initiating a transaction and confirmation of the block containing
that transaction. Categorizing transactions as high (H) and low (L)
based on added fees, their findings indicate that the average confirmation
time for L-class transactions surpasses that of H-class transactions.
Additionally, they noted that when the arrival rate of low-fee transactions
increases fourfold compared to the current rate, these transactions
endure notably large confirmation times within a 1 MB maximum block
size scenario. Utilizing $f(\lambda)$ as a function of lambda for
$E[T]$, and $E[T_{H}]$ and $E[T_{L}]$ representing transaction
confirmation time for high and low priority transactions, they derived
equations \citep{65}:
\begin{equation}
E[T_{H}]=f(\lambda_{H})
\end{equation}
\begin{equation}
E[T_{L}]=\lambda+1f(\lambda_{H}+\lambda_{L})-\lambda f(\lambda_{H}).
\end{equation}

Through validation via discrete-event simulation, the authors showed
that the block generation time in Bitcoin follows an exponential distribution.
Their examination of transaction confirmation times for H and L transactions
concluded that enlarging the maximum block size does not fundamentally
address transaction confirmation time and scalability concerns.

In 2017, Kawase and Kasahara \citep{64} explored Bitcoin's transaction
confirmation time, employing queueing theory as a framework. They
modeled Bitcoin mining as an $M/G^{B}/1$ queue with batch service.
The model assumed that transaction arrivals followed a Poisson distribution,
while service time intervals were considered to have a general distribution.
Within the proposed model, incoming transactions couldn't get immediate
service, even if there was room in the service unit, below the block
size limit (the maximum batch defined). Their research encompassed
numerical analyses of service time and transaction confirmation time-
sojourn time- with a focus on how block size or batch size influenced
these metrics. The authors also conducted a Monte Carlo simulation
to validate their analytical results. Comparing these results to measured
data, they observed close alignment, with a relative error of 1.35\%.
Real-world data from a 2-year period from blockchain.info \citep{86}
informed network simulations, revealing variations between analytical
and simulated results for small block sizes but strong agreement as
block sizes increased. Their work delved into the complexities of
joint distribution and differential-difference equations but left
a unique solution to future research.

In 2019, Ricci et al. \citep{63} presented a framework that combines
machine learning and queueing theory to analyze Bitcoin transaction
confirmation times. This framework incorporates several key factors
into the queueing model, including Activity time, Mean time between
transactions, Mean time between block confirmations, Number of active
blocks, and Transaction delay. To determine which transactions are
likely to be confirmed, they employed machine learning techniques
to classify transactions as confirmed or not. The framework introduces
a threshold, below which it's advisable to wait for natural validation.
If the transaction gets confirmed by the expected time, it's considered
an early confirmation. If not, a classifier is utilized to predict
whether the transaction will ultimately be confirmed, taking into
account the time it has already spent in the system. Additionally,
the researchers employed an $M/G/1$ queueing model to investigate
factors associated with transaction confirmation time. They found
that there was a probability $p$ that the users might need to wait
for an additional block for final confirmation. According to \citep{63}
transaction delay $D$ can be expressed as:
\begin{equation}
E(D)=pE(B)+E(B_{r})
\end{equation}

which $B$ is the time interval between block confirmations and $B_{r}$
denotes the residual time of the inter-block time. Ricci et al.'s
findings using data measurements across more than 3000 blocks in 2016
indicated that unconfirmed transactions typically had lower average
fees compared to confirmed ones. Interestingly, the size of transactions
did not influence their likelihood of confirmation. However, the probability
of confirmation was positively correlated with the transaction value;
larger transactions had a higher chance of being confirmed. Approximately
90\% of all transactions were confirmed within 33 minutes, but typical
users experienced delays exceeding the time between block generations.

In \citep{69} and \citep{67}, Misic et al. introduced a comprehensive
analytical model aimed at understanding Bitcoin's distribution network.
They employed branching processes to reflect random node connectivity.
Further, the authors applied the Jackson network model, treating individual
nodes as priority M/G/1 queuing systems and characterizing data arrival
as a non-homogeneous Poisson process derived from the data delivery
protocol's analytical model. While presenting performance results,
the study highlights various metrics such as block and transaction
distribution times, node response times, forking probabilities, network
partition sizes, and ledger inconsistency period duration.

Zhao et al. \citep{83} expanded the traditional blockchain queue
model by considering a potential zero-transaction scenario. They introduced
a M/G/1 queueing model with limited batch service to replicate how
transactions get confirmed. This model assumed that transactions arrived
randomly, following a Poisson distribution. In their model, they likened
the mining process to a vacation, and the block verification process
to a service. The time taken for mining (V) and block verification
(S) were both i.i.d and followed general distributions. The authors
derived the average number of transactions in the system and the average
confirmation time of transactions using this model. The study concluded
by validating their model using both analysis and simulations, assessing
how well it portrayed the blockchain system's performance.

In Bitcoin, the transactions are stored in a transaction queue named
Mempool to await processing. Miners select transactions from this
queue to form new blocks. Ramezan et al. \citep{62} conducted a study
to investigate the miner's transaction selection process and its impact
on the average waiting time for transactions. In the proposed $M/G/1$
queueing model depicted in Fig 3 , transaction arrivals are denoted
as $\lambda$, following a Poisson process. The average time required
for mining has an exponential distribution and is denoted by $\mu$,
which remains independent of the number of users. 

This paper introduces a novel mining strategy where the miner waits
for D transactions to accumulate in the queue before initiating a
new round of mining. This new approach is labeled $Wait-Min(D)$,
in contrast to the traditional method known as $Wait-Min(1)$. Denoting
the average queue length by $L(D)$ and utilizing Little's law, the
average waiting time per transaction can be calculated as $\overline{W}(D)=\frac{\overline{L}(D)}{\lambda}$.
The authors in \citep{62} derived the formula for the average waiting
time per transaction in traditional mining as:

\begin{equation}
\overline{W}(1)=\frac{\lambda(\lambda+\mu)}{\mu(\lambda\mu+\mu^{2}+\lambda^{2})}
\end{equation}
To validate the newly proposed strategy, the authors conducted simulations
and compared the results obtained with their analytical findings to
assess the performance of the new method. Their investigation revealed
that the optimal value for D, which minimizes the average waiting
time, is approximately $0.9\lambda/\mu$. The results showed that
the average transaction waiting time can be reduced by 10\% when the
miner does not immediately start the next mining process, using all
the stored transactions in the queue.
\begin{figure}[h]
\vspace{-.2cm} 

\includegraphics[scale=0.45]{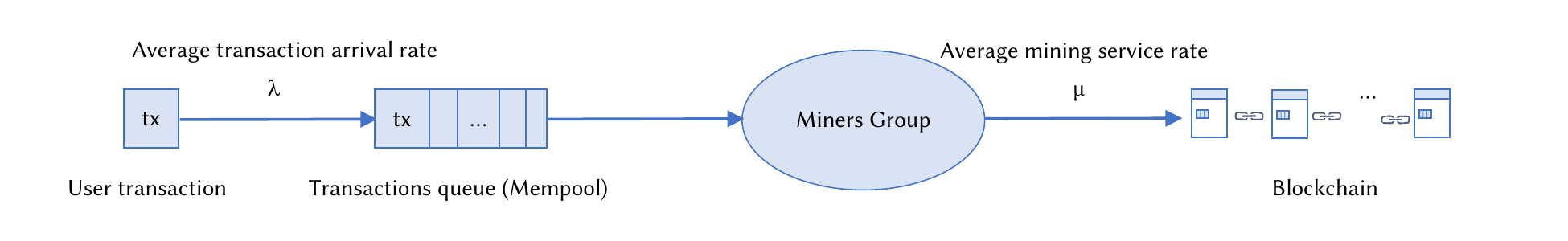}

\vspace{-.3cm} \caption{The M/G/1 queueing model proposed in \citep{62}}

\vspace{-.2cm} 
\end{figure}

Fang and Liu \citep{70} proposed a queueing analytical model to allocate
mining resources in PoW-based blockchain networks, utilizing Lyapunov
optimization techniques and emphasizing a tradeoff between mining
energy and queueing delay. Notably, the proposed model avoided statistical
assumptions on arrivals and services, ensuring a dynamic allocation
algorithm without requiring knowledge of arrival probability distributions
or prioritization mechanisms. The proposed algorithm in this paper,
parameterized by $K$, balances mining energy cost and queueing delay,
showcasing an $O(1/K)$ proximity to optimality and a worst-case delay
scaling as $O(K)$. 

In 2021, Kasahara \citep{66} analyzed three queuing models including
$M/G^{b}/1$ model, $M/G^{b}/1$ model with priority discipline, and
$G/M^{b}/1$ to find the most accurate approach to model the Bitcoin
transaction confirmation process. Kasahara found that the $M/G^{b}/1$
model without transaction priority yielded a transaction confirmation
time half of the measured value, suggesting that arriving transactions
were not included in the mining block. The $M/G^{b}/1$ model with
transaction priority, classifying transactions into Low (L) and High
(H) priority, revealed a non-work-conserving queuing mechanism, indicating
intentional oversight of low-fee transactions by miners. Lastly, the
$G/M^{b}/1$ queue model, with transaction interarrival times following
a general distribution and block generation time following an exponential
distribution with rate $\lyxmathsym{\textmu}$, showed no significant
difference between measured and estimated values using the EM algorithm
\citep{87}. Data-driven simulation results exhibited a significant
transaction confirmation time difference with a small block size,
diminishing as the block size increased, highlighting the third queuing
model provided more accurate transaction confirmation time results
under the current block size of 1 MB.

\subsubsection{Modeling used M/M/1 queue.}

In 2019, Srivastava \citep{75} employed an $M/M/1$ queueing model
to examine transaction confirmation time and applied a Markov queueing
model to regulate block approval rates within the blockchain. The
model's accuracy was verified through experimental outcomes. Using
this model, Srivastava derived the following formula for determining
the average waiting time for a block to be added to the blockchain
at time t, W(t), in which $\lambda$ is the block arrival rate and
$\lyxmathsym{\textmu}$ is the block departure rate \citep{75}.

\begin{equation}
W(t)=\frac{1}{\mu}\left[\frac{\lambda}{\mu}\left(1-\frac{\lambda}{\mu}\right)\right]
\end{equation}

In \citep{71}, the authors presented a queuing theory-based model
to enhance understanding of blockchain performance. Performance parameters,
including the number of transactions in a block, Mining time, Throughput,
Memory pool count, Waiting time in the memory pool, the number of
unconfirmed transactions, total transactions, and generated blocks,
were systematically examined. In the model, the Memory pool, housing
unconfirmed transactions, is represented as an M/M/1 queue, and the
miners' pool, composed of miners collaborating to solve mining puzzles,
is modeled as an M/M/c queue with multiple servers. The policy of
queue is assumed as FCFS, and transaction arrival rates follow a Poisson
distribution. The model is depicted in Fig 4.
\begin{figure}[h]
\vspace{-.2cm} 

\includegraphics[scale=0.5]{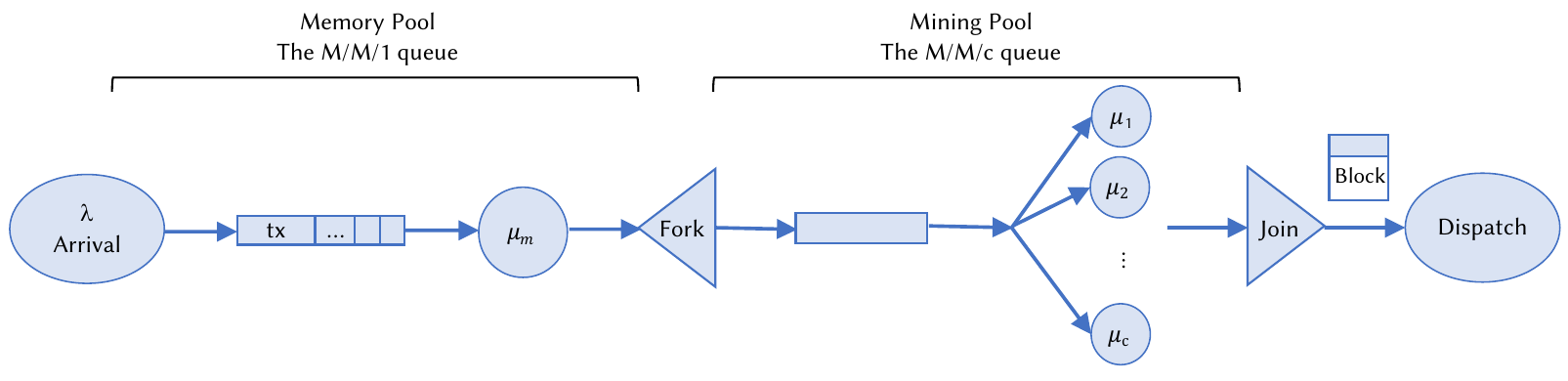}

\vspace{-.3cm} \caption{The proposed model for blockchain systems in \citep{71}}
\vspace{-.3cm} 
\end{figure}

Cao et al. \citep{89} investigated the performance of PoW, PoS, and
DAG consensus algorithms by analyzing Block generation time, Confirmation
delay, Transaction per second, and Confirmation failure probability.
It assumed transaction arrivals follow a Poisson distribution and
mining times as service times follow an exponential distribution.
The study revealed that network resources such as computation power,
coin age, and queue size, along with network load conditions like
transaction arrival rate, significantly impact consensus algorithm
performance. PoW and PoS are more sensitive to network resource parameters,
while DAG is more influenced by network load conditions. The authors
derived the following formulas for the mentioned parameters, assuming
$D$ as difficulty, $U$ as a nonce target, $r_{i}$ as the hash power
of node i, $\lambda$ as transaction arrival rate, $L$ as block size,
$Q_{i}$ as queue length at node i, and $m$ as the number of cumulative
blocks required to confirm a transaction \citep{89}.

\begin{table}[h]
\caption{The formulas presented in \citep{89} for the performance parameters
of PoW and PoS-based blockchains}
\vspace{-.2cm} 

\begin{tabular*}{1\linewidth}{@{\extracolsep{\fill}}lll}
\toprule 
\multirow{1}{*}{{\footnotesize{}Parameter}} & \multirow{1}{*}{{\footnotesize{}Equation for PoW}} & {\footnotesize{}Equation for PoS}\tabularnewline
\midrule
\midrule 
\multirow{2}{*}{{\footnotesize{}Block Time}} & \multirow{2}{*}{{\footnotesize{}$T_{b}^{pow}=\frac{D}{\sum_{i=1}^{n}r_{i}}$}} & \multirow{2}{*}{{\footnotesize{}$T_{b}^{pos}=\frac{D}{\sum_{i=1}^{n}bal_{i}\times t_{i}}$}}\tabularnewline
 &  & \tabularnewline
\midrule
\multirow{2}{*}{{\footnotesize{}Confirmation delay}} & \multirow{2}{*}{{\footnotesize{}$T_{c}^{pow}=m^{pow}\times T_{b}^{pow}$}} & \multirow{2}{*}{{\footnotesize{}$T_{c}^{pos}=m^{pos}\times T_{b}^{pos}$}}\tabularnewline
 &  & \tabularnewline
\midrule 
\multirow{4}{*}{{\footnotesize{}TPS}} & \multirow{4}{*}{{\footnotesize{}$TPS^{pow}=\begin{cases}
\sum_{i=1}^{n}\lambda_{i} & \sum_{i=1}^{n}\lambda_{i}\leq\frac{L}{T_{b}^{pow}}\\
\frac{L}{T_{b}^{pow}} & \sum_{i=1}^{n}\lambda_{i}>\frac{L}{T_{b}^{pow}}
\end{cases}$}} & \multirow{4}{*}{{\footnotesize{}$TPS^{pos}=\begin{cases}
\sum_{i=1}^{n}\lambda_{i} & \sum_{i=1}^{n}\lambda_{i}\leq\frac{L}{T_{b}^{pos}}\\
\frac{L}{T_{b}^{pos}} & \sum_{i=1}^{n}\lambda_{i}>\frac{L}{T_{b}^{pos}}
\end{cases}$}}\tabularnewline
 &  & \tabularnewline
 &  & \tabularnewline
 &  & \tabularnewline
\bottomrule
\end{tabular*}

\vspace{-.1cm} 
\end{table}

To explore block broadcasting in wireless networks and evaluate the
performance of blockchain radio access networks (B-RAN) \citep{1},
Ling et al. \citep{74} employed a time-homogeneous Markov chain to
establish an M/M/1 queueing model. They segmented the service process
of a valid request in B-RAN into four stages: 1) waiting for inclusion
into a block; 2) waiting for confirmations; 3) waiting for service;
and 4) being in service. The authors established multiple queues to
model these phases. However, they noted that during the third stage,
requests within the same block arrive simultaneously, and the number
of requests correlates with the block generation time. Consequently,
this queue becomes non-Markovian because prior events, apart from
the current queue state, may influence its future state. Furthermore,
Ling et al. derived steady-state probabilities for various latency-based
metrics. By validating the proposed model with experimental data,
they demonstrated its potential to offer valuable insights for future
designs of B-RAN.

To address the question of whether increasing the block size improves
transaction waiting time in Bitcoin, Vesely et al. \citep{72} used
an M/M/1 queue model to scrutinize the transaction mining process.
This proposed model encompasses two stages including the \textquotedbl Block
generation\textquotedbl{} stage and the \textquotedbl Blockchain
building\textquotedbl{} stage similar to the mode proposed in \citep{76}.
The model is employed to compare the number of transactions awaiting
processing in two scenarios: a block size of 1 MB and a block size
of 5 MB. Through this comparison, the authors demonstrated that enlarging
the block size from 1 MB to 5 MB decelerates the growth of the transaction
queue length in Bitcoin. Nevertheless, this enlargement in block size
results in reduced server utilization in Bitcoin and poses challenges
for users dealing with the storage of a larger blockchain.

\subsubsection{Modeling used G/M/1 queue.}

The authors in \citep{78} employed a G/M/1 queueing model with batch
service and general input to comprehend the stochastic nature of the
transaction confirmation process. They aimed to predict transaction-confirmation
times by comparing analytical outcomes with trace-driven simulations
in numerical examples. The findings indicated that, for the current
maximum block size limit of 1 MB, exponential distributions offered
the most precise estimation of the mean transaction confirmation time.
Moreover, the study explored the relationship between transaction
arrival rates, block size, and transaction confirmation times. Enhancing
the block size has the potential to incrementally boost transaction
throughput (TPS); however, this improvement doesn't fundamentally
resolve scalability challenges. Additionally, it was noted that while
the transaction confirmation time gradually increased with the coefficient
of variation of exponential distribution, this effect became insignificant
with larger block sizes.

In 2018, Li et al. \citep{76} constructed a Markovian batch-service
queueing model with two stages to analyze transaction confirmation
processes in blockchains. Utilizing the matrix geometric solution,
they identified stability conditions for the system and assessed three
key performance metrics: 1) average queued transactions, 2) average
transactions within a block, and 3) average transaction confirmation
time. Their proposed GI/M/1 queueing model features two stages: the
\textquotedbl Block generation\textquotedbl{} stage with a service
rate of \textgreek{m}2 and the \textquotedbl Blockchain building\textquotedbl{}
stage with a service rate of \textgreek{m}1. Both mentioned times
are i.i.d distributed, following an exponential distribution. Similarly,
blockchain building times in stage 2 follow an i.i.d exponential distribution
with a service rate \textgreek{m}1. The model also assumes transactions
are processed on an (FCFS) basis from the transaction pool.

Results indicated that the average transaction confirmation time decreases
as the block size increases. Furthermore, they observed a threshold
value \textgreek{h}, where for b \ensuremath{\le} \textgreek{h}, E{[}T{]}
increases as \textgreek{l} increases, whereas for b > \textgreek{h},
E{[}T{]} increases as \textgreek{l} decreases. Although based on simplistic
exponential or Poisson assumptions, this model suggests future avenues
for research in blockchain queueing theory. The queueing model proposed
in \citep{76} can be observed in Fig 5.
\begin{figure}[h]
\vspace{-.4cm} 

\includegraphics[scale=0.45]{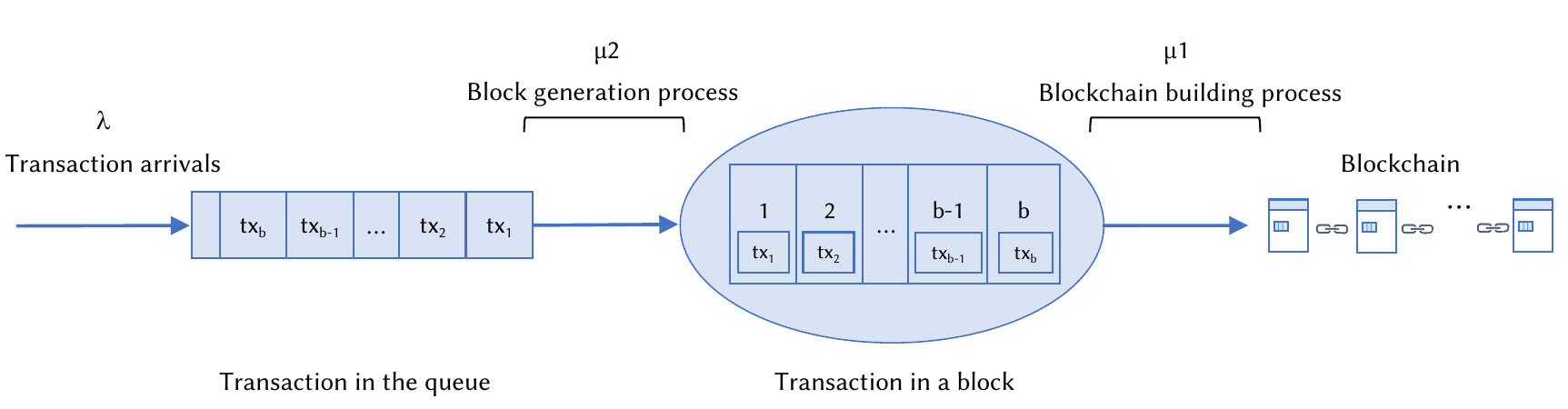}

\vspace{-.3cm} \caption{The Blockchain queueing system proposed in \citep{76}}
\vspace{-.4cm} 
\end{figure}
One year later, Li et al. extended the queueing model introduced in
\citep{77} by making several modifications. They altered the service
discipline from (FCFS) to service-in-random-order, transitioned from
exponential to phase-type (PH) service times---which included both
blockchain generation and building--- and shifted from Poisson to
Markov arrival process (MAP) for transaction arrivals. Initially,
they established a stable condition for the blockchain system utilizing
the matrix-geometric solution. To address the challenges in calculating
transaction confirmation times in \citep{77}, the authors devised
a computational technique for the first passage times, employing both
infinite-sized phase-type distributions and RG factorizations. Additionally,
Li et al. delved into some key performance metrics, such as the average
number of transactions in the queueing waiting room, the average number
of transactions within a block, and the average transaction confirmation
time. Finally, they validated the computational feasibility of the
presented theoretical findings through numerical examples.

\subsubsection{Modeling used GI/GI/1 queue.}

Geissler et al. \citep{79} constructed a discrete-time $GI/GI^{N}/1$
queueing model aimed at analyzing queue size and transaction waiting
times within Proof-based blockchains. This model facilitated the prediction
and assessment of queue size and waiting time distributions, considering
diverse input parameters. In an effort to simplify and generalize
the model, the authors made several assumptions, including an infinite
transaction pool, a discrete-time block generation process, a single
server without propagation delays among network nodes, and fixed service
times. Nonetheless, the proposed model deviates from real-world processes
and lacks the required accuracy and practical applicability.

\subsubsection{Modeling used other queue types.}

Frolkova et al. \citep{80} undertook an examination of busy periods'
distribution and service delays within the Bitcoin network. They approached
the network's synchronization process by formulating it as an infinite
server $G/M/\infty$ model with batch departures. In their study,
the authors managed to derive an analytical solution for the model,
enabling the capture of the queue's stationary distribution. Additionally,
they introduced a random-style fluid limit considering service latencies.
The authors utilized a connection between two particular service policies
to acquire the steady-state distribution for the model.

Expanding on Frolkova's model, Fralix \citep{81} introduced an additional
perspective on modeling the blockchain transaction confirmation process
as an infinite-server queue. Fralix demonstrated that a variant of
the distributional Little's law, which is time-dependent, serves as
an effective tool to investigate the time-varying characteristics
inherent within this model.

Lian et al. \citep{82} devised a stock trading model utilizing an
$M/M/N/m$priority-based queue model characterized by $N$ channels
and m users, inclusive of three priority classes with arrival rates:
$\lambda_{1}$, $\lambda_{2}^{k=1}$, and $\lambda_{2}^{k=2}$. The
model comprises two phases: the block generation phase and the chain
establishment phase. In the block generation phase, incoming data
flows queue within an infinite waiting room and are categorized into
high-priority and low-priority streams, awaiting assembly into blocks.
Subsequently, in the chain establishment phase, newly formed blocks
are concatenated to form a blockchain. Each consensus mechanism operates
within different delays, which the study does not explicitly discuss.
These stages facilitate the derivation of average waiting times and
queue lengths for various priority data streams within a blockchain.
The proposed queueing model underwent simulation and measurement,
assessing throughput, delay, and channel utilization, and compared
with an FCFS-based queuing model.

\subsection{Performance modeling using other modeling approaches}

In their pioneering work detailed in \citep{88}, the authors extended
the Markov Decision Process (MDP) to create a comprehensive framework
for evaluating the security and performance of blockchains. Additionally,
they integrated an MDP security model with a Bitcoin simulator to
form a unified framework, facilitating the assessment of blockchain
security by utilizing the simulator's stale block rate as input for
the MDP model. However, it's important to note that the framework
primarily concentrates on propagation delay and does not address the
crucial processing delay inherent in solving consensus puzzles within
PoW-based systems. Furthermore, the framework does not encompass queueing
delay. The proposed framework depicted in \citep{88} is illustrated
in Fig 6.
\begin{figure}
\includegraphics[scale=0.45]{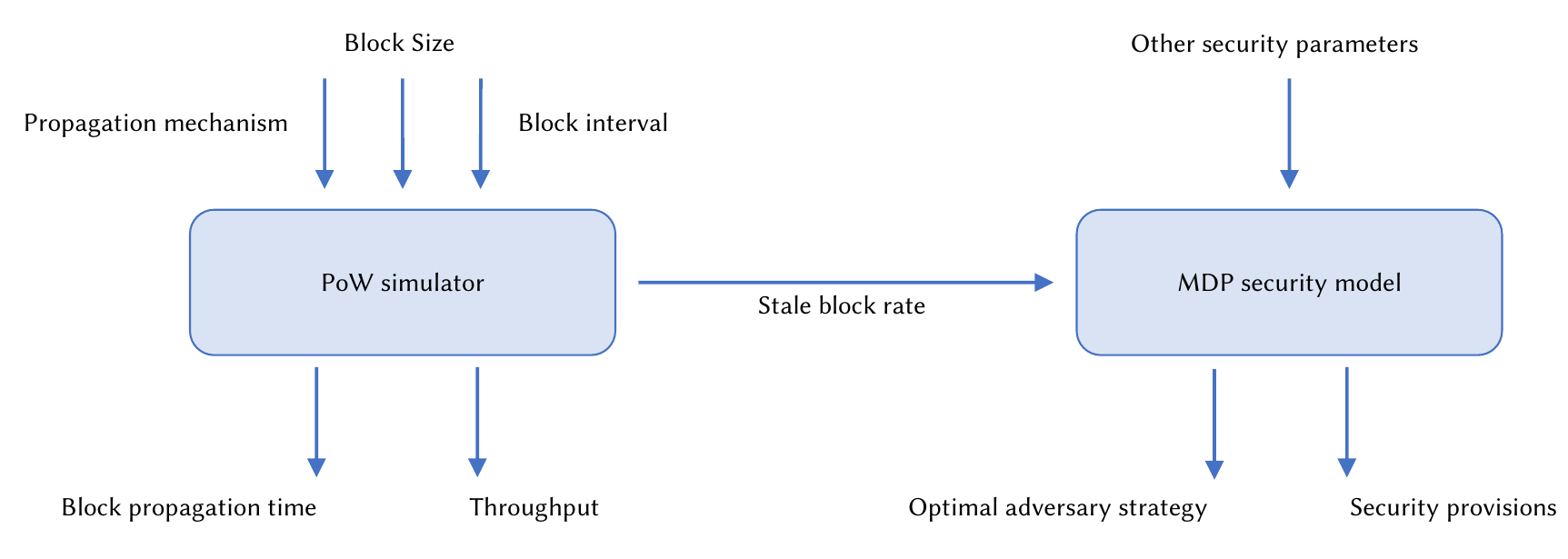}\vspace{-.3cm} 

\caption{The structure of the proposed framework in \citep{88}}
\vspace{-.5cm} 
\end{figure}

Zhang et al. \citep{85} introduced a predictive model based on Ethereum's
\textquotedbl World State\textquotedbl{} structure to anticipate
contract execution performance and storage requirements based on transaction
volume. Additionally, the research delved into Ethereum's \textquotedbl World
State\textquotedbl{} and its Merkle Patricia Tree (MPT) implementation
and analyzed its core impact on performance and storage. Through an
examination of the Modified Merkle Patricia Tree (MPT), the study
establishes a logarithmic relationship between MPT performance/storage
and transaction volume, offering valuable insights for businesses
in decision-making and early warnings regarding system performance
and storage needs. By comparing experimental data with predictions,
the study demonstrates the model's relative accuracy.

A random network is characterized by the probability of an edge existing
between two nodes, which is not influenced by the nodes' characteristics.
Essentially, the degree distribution $P(k)$ becomes the sole relevant
function in this scenario \citep{90}. Two widely used models in various
applications are $G(n,m)$ proposed by Erd\H{o}s and R�nyi \citep{92}
and $G(n,p)$ recommended by Edgar Gilbert \citep{93}. In the $G(n,m)$model,
all graphs consisting of n nodes and m edges are assigned equal probabilities,
whereas in the $G(n,p)$ model, each potential edge among n nodes
emerges independently with a constant probability p \citep{91}.

In 2019, Shahsavari et al. \citep{84} introduced an analytical model
utilizing the Erdos-Renyi random graph \citep{92} to model performance
and analyze the inventory-based protocol for block dissemination in
Bitcoin. Validating their model using OMNet++ network simulator \citep{94},
they compared its accuracy using data extracted from the Bitcoin network.
The findings highlighted a tradeoff among the default connections
per node, network bandwidth, and block size, crucial in determining
the optimal block propagation delay across the network. A comprehensive
summary of performance modeling studies regarding public permissionless
blockchains is provided in Table 4.
\begin{table}
\caption{Summary of performance models for public permissionless blockchain
systems}
\vspace{-.2cm} 

\begin{tabular*}{1\linewidth}{@{\extracolsep{\fill}}clllll}
\toprule 
 & {\footnotesize{}Model} & {\footnotesize{}Ref} & {\footnotesize{}Platform} & {\footnotesize{}Performance metrics} & {\footnotesize{}Validation}\tabularnewline
\midrule
\midrule 
\multirow{26}{*}{\begin{turn}{90}
{\footnotesize{}Queueing theory}
\end{turn}} & {\scriptsize{}M/G/1} & {\scriptsize{}\citep{62}} & {\scriptsize{}BTC} & {\scriptsize{}Tx average waiting time} & {\scriptsize{}Simulation}\tabularnewline
\cmidrule{3-6} \cmidrule{4-6} \cmidrule{5-6} \cmidrule{6-6} 
 &  & {\scriptsize{}\citep{63}} & {\scriptsize{}BTC} & {\scriptsize{}Tx confirmation time/ Tx delay} & {\scriptsize{}Measured data}\tabularnewline
\cmidrule{3-6} \cmidrule{4-6} \cmidrule{5-6} \cmidrule{6-6} 
 &  & {\scriptsize{}\citep{64}} & {\scriptsize{}BTC} & {\scriptsize{}Tx sojourn (Response time)} & {\scriptsize{}Simulation}\tabularnewline
\cmidrule{3-6} \cmidrule{4-6} \cmidrule{5-6} \cmidrule{6-6} 
 &  & {\scriptsize{}\citep{65}} & {\scriptsize{}BTC} & {\scriptsize{}Mean confirmation time} & {\scriptsize{}Numerical analysis}\tabularnewline
\cmidrule{3-6} \cmidrule{4-6} \cmidrule{5-6} \cmidrule{6-6} 
 &  & {\scriptsize{}\citep{66}} & {\scriptsize{}BTC} & {\scriptsize{}Tx confirmation time} & {\scriptsize{}Measured data}\tabularnewline
\cmidrule{3-6} \cmidrule{4-6} \cmidrule{5-6} \cmidrule{6-6} 
 &  & {\scriptsize{}\citep{67}} & {\scriptsize{}BTC} & {\scriptsize{}Tx rate/ Block rate/ Node response time} & {\scriptsize{}Experimentation}\tabularnewline
\cmidrule{3-6} \cmidrule{4-6} \cmidrule{5-6} \cmidrule{6-6} 
 &  & {\scriptsize{}\citep{69}} & {\scriptsize{}BTC} & {\scriptsize{}Node response time} & {\scriptsize{}Experimentation}\tabularnewline
\cmidrule{3-6} \cmidrule{4-6} \cmidrule{5-6} \cmidrule{6-6} 
 &  & {\scriptsize{}\citep{83}} & {\scriptsize{}BTC} & {\scriptsize{}Tx confirmation time/ Mean number of txs} & {\scriptsize{}Simulation}\tabularnewline
\cmidrule{3-6} \cmidrule{4-6} \cmidrule{5-6} \cmidrule{6-6} 
 &  & {\scriptsize{}\citep{70}} & {\scriptsize{}PoW} & {\scriptsize{}Queue delay} & {\scriptsize{}Simulation}\tabularnewline
\cmidrule{3-6} \cmidrule{4-6} \cmidrule{5-6} \cmidrule{6-6} 
 &  & {\scriptsize{}\citep{176}} & {\scriptsize{}BTC} & {\scriptsize{}Average response time} & {\scriptsize{}Simulation}\tabularnewline
\cmidrule{2-6} \cmidrule{3-6} \cmidrule{4-6} \cmidrule{5-6} \cmidrule{6-6} 
 & {\scriptsize{}M/M/1} & {\scriptsize{}\citep{71}} & {\scriptsize{}BTC/ETH} & {\scriptsize{}Number of txs per block/Throughput/ Mempool size/ tx
waiting time} & {\scriptsize{}Actual data}\tabularnewline
\cmidrule{3-6} \cmidrule{4-6} \cmidrule{5-6} \cmidrule{6-6} 
 &  & {\scriptsize{}\citep{72}} & {\scriptsize{}BTC} & {\scriptsize{}The number of waiting txs in queue} & {\scriptsize{}Actual data}\tabularnewline
\cmidrule{3-6} \cmidrule{4-6} \cmidrule{5-6} \cmidrule{6-6} 
 &  & {\scriptsize{}\citep{73}} & {\scriptsize{}ETH} & {\scriptsize{}Tx confirmation time/ Block throughput} & {\scriptsize{}Simulation}\tabularnewline
\cmidrule{3-6} \cmidrule{4-6} \cmidrule{5-6} \cmidrule{6-6} 
 &  & {\scriptsize{}\citep{74}} & {\scriptsize{}B-RAN} & {\scriptsize{}Latency} & {\scriptsize{}Simulation}\tabularnewline
\cmidrule{3-6} \cmidrule{4-6} \cmidrule{5-6} \cmidrule{6-6} 
 &  & {\scriptsize{}\citep{75}} & {\scriptsize{}NA} & {\scriptsize{}Waiting time/ Tx confirmation time} & {\scriptsize{}Experimentation}\tabularnewline
\cmidrule{3-6} \cmidrule{4-6} \cmidrule{5-6} \cmidrule{6-6} 
 &  & {\scriptsize{}\citep{171}} & {\scriptsize{}BTC} & {\scriptsize{}Average number of transactions/ Average confirmation
time} & {\scriptsize{}Numerical analysis}\tabularnewline
\cmidrule{3-6} \cmidrule{4-6} \cmidrule{5-6} \cmidrule{6-6} 
 &  & {\scriptsize{}\citep{173}} & {\scriptsize{}ETH} & {\scriptsize{}Average waiting time/ Throughput} & {\scriptsize{}Simulation}\tabularnewline
\cmidrule{3-6} \cmidrule{4-6} \cmidrule{5-6} \cmidrule{6-6} 
 &  & {\scriptsize{}\citep{89}} & {\scriptsize{}BTC/ETH} & {\scriptsize{}Block generation time/Confirmation delay/TPS/Failure
probability} & {\scriptsize{}Simulation}\tabularnewline
\cmidrule{3-6} \cmidrule{4-6} \cmidrule{5-6} \cmidrule{6-6} 
 &  & {\scriptsize{}\citep{175}} & {\scriptsize{}BTC/ETH} & {\scriptsize{}Transaction queue size/ Block waiting time} & {\scriptsize{}Simulation}\tabularnewline
\cmidrule{2-6} \cmidrule{3-6} \cmidrule{4-6} \cmidrule{5-6} \cmidrule{6-6} 
 & {\scriptsize{}G/M/1} & {\scriptsize{}\citep{76}} & {\scriptsize{}BTC} & {\scriptsize{}Average tx confirmation time/ Average number of txs
in queue and a block} & {\scriptsize{}Numerical analysis}\tabularnewline
\cmidrule{3-6} \cmidrule{4-6} \cmidrule{5-6} \cmidrule{6-6} 
 &  & {\scriptsize{}\citep{77}} & {\scriptsize{}BTC} & {\scriptsize{}Mean stationary number of txs in queue and block} & {\scriptsize{}Numerical analysis}\tabularnewline
\cmidrule{3-6} \cmidrule{4-6} \cmidrule{5-6} \cmidrule{6-6} 
 &  & {\scriptsize{}\citep{78}} & {\scriptsize{}BTC} & {\scriptsize{}Tx confirmation time} & {\scriptsize{}Simulation}\tabularnewline
\cmidrule{2-6} \cmidrule{3-6} \cmidrule{4-6} \cmidrule{5-6} \cmidrule{6-6} 
 & {\scriptsize{}GI/GI/1} & {\scriptsize{}\citep{79}} & {\scriptsize{}ETH} & {\scriptsize{}Tx waiting time/ Queue size} & {\scriptsize{}Measured data}\tabularnewline
\cmidrule{2-6} \cmidrule{3-6} \cmidrule{4-6} \cmidrule{5-6} \cmidrule{6-6} 
 & {\scriptsize{}GI/M/$\infty$} & {\scriptsize{}\citep{80}} & {\scriptsize{}BTC} & {\scriptsize{}Queue length} & {\scriptsize{}Numerical analysis}\tabularnewline
\cmidrule{3-6} \cmidrule{4-6} \cmidrule{5-6} \cmidrule{6-6} 
 &  & {\scriptsize{}\citep{81}} & {\scriptsize{}BTC} & {\scriptsize{}Time-dependent behavior} & {\scriptsize{}Numerical analysis}\tabularnewline
\cmidrule{2-6} \cmidrule{3-6} \cmidrule{4-6} \cmidrule{5-6} \cmidrule{6-6} 
 & {\scriptsize{}M/M/N} & {\scriptsize{}\citep{82}} & {\scriptsize{}NA} & {\scriptsize{}Tx delay/ Channel utilization} & {\scriptsize{}Experimentation}\tabularnewline
\midrule 
\multirow{7}{*}{\begin{turn}{90}
{\footnotesize{}Other models}
\end{turn}} & {\scriptsize{}Prediction model} & {\scriptsize{}\citep{85}} & {\scriptsize{}ETH} & {\scriptsize{}Executing contract's Performance and storage prediction} & {\scriptsize{}Experimentation}\tabularnewline
\cmidrule{2-6} \cmidrule{3-6} \cmidrule{4-6} \cmidrule{5-6} \cmidrule{6-6} 
 & {\scriptsize{}Stochastic networks} & {\scriptsize{}\citep{68}} & {\scriptsize{}ETH} & {\scriptsize{}Block generation rate} & {\scriptsize{}Simulation}\tabularnewline
\cmidrule{2-6} \cmidrule{3-6} \cmidrule{4-6} \cmidrule{5-6} \cmidrule{6-6} 
 & {\scriptsize{}Random graph model} & {\scriptsize{}\citep{84}} & {\scriptsize{}BTC} & {\scriptsize{}Block propagation delay/ Traffic workload} & {\scriptsize{}Experimentation}\tabularnewline
\cmidrule{2-6} \cmidrule{3-6} \cmidrule{4-6} \cmidrule{5-6} \cmidrule{6-6} 
 & {\scriptsize{}Markov Decision} & {\scriptsize{}\citep{88}} & {\scriptsize{}BTC} & {\scriptsize{}Transaction per second Block propagation time} & {\scriptsize{}Simulation}\tabularnewline
\cmidrule{3-6} \cmidrule{4-6} \cmidrule{5-6} \cmidrule{6-6} 
 & {\scriptsize{}Process (MDP)} & {\scriptsize{}\citep{172}} & {\scriptsize{}BTC} & {\scriptsize{}Network stability} & {\scriptsize{}Simulation}\tabularnewline
\cmidrule{2-6} \cmidrule{3-6} \cmidrule{4-6} \cmidrule{5-6} \cmidrule{6-6} 
 & {\scriptsize{}Sleep queueing system} & {\scriptsize{}\citep{174}} & {\scriptsize{}BTC} & {\scriptsize{}Average tx confirmation time} & {\scriptsize{}Experimentation}\tabularnewline
\cmidrule{2-6} \cmidrule{3-6} \cmidrule{4-6} \cmidrule{5-6} \cmidrule{6-6} 
 & {\scriptsize{}Game theory} & {\scriptsize{}\citep{177}} & {\scriptsize{}BTC} & {\scriptsize{}Transactions/ waiting time in Mempool} & {\scriptsize{}Numerical analysis}\tabularnewline
\bottomrule
\end{tabular*}
\end{table}
\begin{figure}
\includegraphics[scale=0.46]{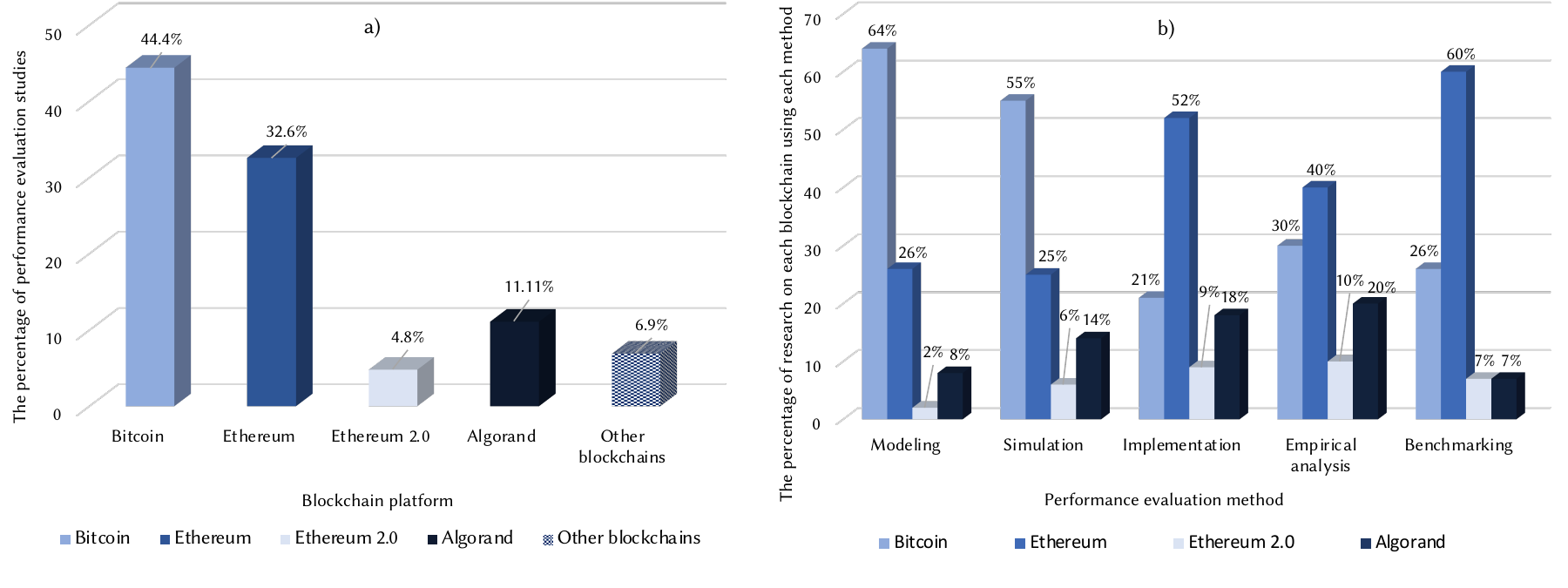}

\caption{a) Research papers published per performance evaluation method and
blockchain platform, b) Research papers published on public permissionless
blockchains per blockchain platform.}

\includegraphics[scale=0.46]{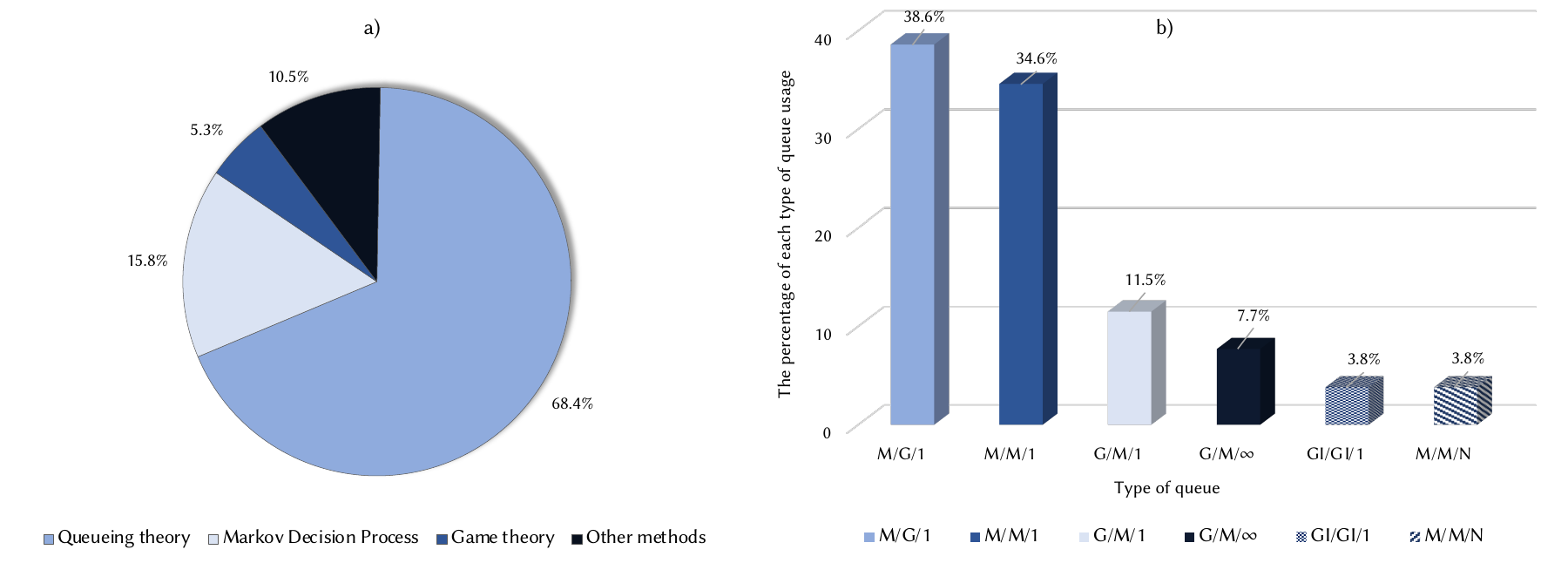}

\caption{a) Research papers published on performance modeling per modeling
method, b) Research papers published on public permissionless blockchains'
performance modeling using queueing theory per type of queue.}
\vspace{-.2cm} 
\end{figure}

\section{Workload and traffic modeling of blockchain systems}

Understanding, analyzing, and modeling traffic and workload holds
significant importance within the realm of performance modeling and
evaluation. A computer system's performance relies heavily on how
it handles network traffic and the workload it receives. For instance,
optimal performance occurs when tasks are evenly distributed, contrasting
with unpredictable bursts that lead to congestion \citep{131}. Therefore,
reliable performance assessments necessitate representative workloads
to yield dependable outcomes. Achieving this involves gathering data
from actual workloads and devising statistical models that encompass
their key attributes. Pioneering studies have delved into modeling,
characterizing, and regenerating internet traffic \citep{127,128,129,130,134}.
The evaluation of performance through workload modeling in computer
networks is extensively discussed in \citep{131,132,133,135,136}.
To explore the specifics of network and traffic modeling within blockchain
networks, we categorized these studies into four groups: 1) Modeling
network topology, 2) Characterizing workloads, 3) Benchmarking via
workload, and 4) Analyzing and modeling traffic. A comprehensive overview
of this section is outlined in Table 5.

\subsection{Network topology modeling}

In 2020, Lao et al. examined existing traffic models in P2P and blockchain
systems and introduced a traffic model specifically tailored for IoT-blockchain
systems \citep{137}. In \citep{138}, Eisenbarth et al. conducted
a study focusing on the public nodes that constitute the structure
of the Bitcoin P2P network.

Numerous studies have concentrated on modeling and analyzing the topology
of blockchain networks. Here, we'll delve into a subset of these papers,
particularly emphasizing studies pertaining to public permissionless
blockchains.

In 2014, Donet et al. \citep{95} conducted a comprehensive data collection,
identifying over 872,000 distinct Bitcoin nodes. This effort aimed
to provide insights into various aspects of the Bitcoin peer-to-peer
network, including its size, geographic distribution of nodes, network
stability, and data related to the transmission speed of information.
In 2015, Miller et al. \citep{96} introduced efficient methods for
mapping the Bitcoin broadcast network and pinpointing nodes with significant
influence within this network. AddressProbe, a technique designed
for uncovering peer-to-peer connections in Bitcoin, is presented in
this paper. Additionally, the authors developed CoinScope, an infrastructure
tailored to facilitate short yet large-scale experiments in Bitcoin,
aimed at supporting AddressProbe and other related tools.

To assess the degree of decentralization in the prominent blockchains
Ethereum and Bitcoin, Gencer et al. \citep{97} conducted a measurement
study involving diverse decentralization metrics for both networks.
They utilized established internet measurement methods and employed
the Falcon Relay Network as a new tool to gather their data. In \citep{98},
Deshpande et al. introduced BTCmap, a framework created specifically
to explore the topology of the Bitcoin network. This framework comprises
two essential modules: a sniffer, which retrieves the local addresses
database from each peer, and a Bitcoin peer emulator, responsible
for selecting neighboring nodes and constructing the topology of the
network. To analyze the underlying peer-to-peer network in Ethereum
blockchain, Kim et al. in \citep{99} created NodeFinder, an innovative
monitoring tool offering an unparalleled perspective of the network
ecosystem in Ethereum. Imtiaz et al. in \citep{100} discovered and
provided experimental evidence for the previously unrecognized impact
of churn---intermittent network connectivity---on the Bitcoin network.
Their study involved a comprehensive analysis of churn, encompassing
the daily churn rate and statistical modeling of session durations
for up and down network connections. This statistical model is valuable
for researchers interested in analyzing, simulating, or replicating
Bitcoin network behavior. Leveraging this statistical characterization,
they assessed how churn influences the block propagation delay within
the live Bitcoin network.

In \citep{101}, the authors introduced a topology discovery system
aimed at creating an accurate graph-based representation of the Bitcoin
network's structure. This system involves live gathering and examination
of Bitcoin's peer-to-peer connections.. They utilized a modified version
of the PageRank algorithm \citep{145}to aggregate incoming node information,
enabling more comprehensive graph analysis. This approach facilitated
insights into various aspects of the Bitcoin network, such as its
size, network stability, and data related to the geographic locations
of Bitcoin nodes. To understand the topological characteristics of
the Ethereum P2P network, Gao et al. \citep{102} performed a measurement
analysis of the Ethereum P2P network. 

Seres et al. in \citep{103} aimed to measure and evaluate the structural
and topological characteristics of Bitcoin's Lightning Network \citep{143},
a method to enhance Bitcoin's scalability. They contended that the
current topological features of Bitcoin's Lightning Network could
be enhanced to improve its security, thus enabling it to achieve its
full potential. In 2020, Lee et al. \citep{104} explored various
local and global graph characteristics across four Ethereum blockchains
(ContractNet, TransactionNet, TraceNet and TokenNet), along with three
notable subnetworks (Binance Coin, Bancor, and Zilliqa). Their investigation
encompassed a comprehensive experimental assessment of these networks.
In \citep{105}the authors delved into analyzing the Bitcoin protocol
using an extended version of the probabilistic model checker PRISM.
By employing probabilistic modeling, the study explored the mechanisms
behind forks and their reliance on specific protocol parameters like
crypto puzzle difficulty and network communication delays. The research
also investigated the behavior of networks with churn miners, those
exiting and later rejoining the network, as well as various network
topologies. Additionally, the study includes simulations of the Bitcoin
protocol, considering diverse network structures to assess whether
these alternative topologies exhibit equal or greater resistance to
forks compared to the original Bitcoin network. 

As the pioneering study exploring the evolution and temporal characteristics
of the Ethereum network, Zhao et al. delved into investigating the
evolutionary patterns, temporal dynamics, and predictive elements
of four Ethereum blockchain interaction networks (TraceNet, ContractNet,
TransactionNet, and TokenNet)\citep{106}. Subsequently, the authors
conducted an extensive empirical assessment of these networks' graphical
representations. In the paper \citep{107}, the authors showcased
several findings derived from their analysis of the Ethereum 2.0 peer-to-peer
network. They obtained this data using their network monitoring tool,
Armiarma, which transparently collected information, providing them
with a comprehensive understanding of the network's current status.
Maeng et al. in \citep{108} presented Search-node, an adapted version
of an Ethereum client for discovering nodes in the Ethereum main network.
They visualized and analyzed the Ethereum network's topology, exploring
various network parameters and delving into node properties, including
relationships, connectivity, geographical distribution, security implications,
and potential attacks targeting influential nodes.

In \citep{109}, the authors introduced a random graph model to analyze
the inventory-based protocol's performance in Bitcoin's block propagation.
By deriving explicit equations for block propagation delay and traffic
overhead, they provided insights into the impact of these parameters
on the network. Additionally, they evaluated the implications of deploying
a relay network and studied how the relay network size affects both
network performance and decentralization. To validate their model,
the authors conducted simulations and compared the results with real
data from the Bitcoin network. In their recent paper, Essaid et al.
\citep{110} conducted a study on the Bitcoin network's behavior and
dynamic topology from 2018 to 2022. They employed Node-Probe, a technique
utilizing recursive scanning to gather snapshots of the Bitcoin main
network. The study focused on analyzing network properties, community
structures, and topology changes. Their findings revealed significant
alterations in the Bitcoin network's characteristics over time, particularly
in the number of active, fresh, and permanent nodes. Furthermore,
their analytical exploration indicated intriguing insights, suggesting
that enhancing the propagation mechanism through master nodes could
potentially reduce propagation delays by approximately 25\% compared
to the default Bitcoin protocol. Zheng et al. \citep{111} pioneered
a study delving into Algorand's network and transactions analysis.
They conducted an examination of Algorand's features using graph analysis.
The team collected external data via an Algorand indexer, establishing
connections to the nodes for the ledger's data processing, followed
by the construction of the graph. Analyzing this graph yielded fresh
observations about this blockchain system.

\subsection{Workload characterization}

Workload characterization is one of the essential field evaluate and
analysis of computer networks. Numerous investigations have concentrated
on workload characterization of networks, like \citep{155,132}. Additionally,
there have been studies delving into specific workloads, such as web
server workload \citep{151,152,153}, web services workload \citep{158,159},
IOT workload \citep{150}, those examined workloads of social networks
\citep{156,157}, more. A comprehensive compilation of these workload
characterization works is available in survey papers within this particular
field \citep{154,147,148}.

The prediction of network behavior within a blockchain system to maintain
essential service benchmarks and network stability demands an understanding
of the network's characteristics and its workload. The study by Spirkina
et al. \citep{112} undertook an examination of Blockchain traffic,
investigating its features in information transmission across the
network. Additionally, the paper surveyed analytical and simulation
modeling solutions, offering an overview that enables the prediction
of network load. In the paper by Vladyko \citep{113}, another investigation
into predicting network behavior within blockchains is conducted.
This study explored methods for modeling these systems and advocates
for the adoption of a simulation system to evaluate network performance
and its components. Furthermore, the paper shared findings from comparative
simulations. A master thesis authored by Jidrot \citep{114} applied
Blockchain technology to securely gather and preserve data from distributed
cloud resources, emphasizing the necessity for data integrity over
extended periods. The primary objectives included implementing a Blockchain
for distributed cloud monitoring, characterizing the workload, and
evaluating the Blockchain system's performance. The thesis defines
an initial system model, develops an implementation based on HLF blockchain,
and conducts experiments to comprehend how design factors and system
inputs influence the system's capabilities and performance. Gebraselase
et al. \citep{126} conducted an extensive analysis of Bitcoin's transaction
characteristics within the Bitcoin blockchain domain. They also provided
an analysis by examining collected data from a measurement setup,
which encompassed blocks and transactions information.
\begin{table}
\caption{Studies on network and traffic modeling in blockchain systems}
\vspace{-.2cm} 

\begin{tabular*}{1\linewidth}{@{\extracolsep{\fill}}llllll}
\toprule 
\textbf{\footnotesize{}Category} & \textbf{\footnotesize{}Platform} & \textbf{\footnotesize{}Ref} & \textbf{\footnotesize{}PP{*}} & \textbf{\footnotesize{}Validation} & \textbf{\footnotesize{}Description}\tabularnewline
\midrule
\midrule 
{\scriptsize{}Network} & {\tiny{}BTC} & {\scriptsize{}\citep{95}} & {\tiny{}\Checkmark{}} & {\scriptsize{}Measurement} & {\scriptsize{}Model and analyze various aspects of the Bitcoin p2p
network}\tabularnewline
\cmidrule{3-6} \cmidrule{4-6} \cmidrule{5-6} \cmidrule{6-6} 
{\scriptsize{}topology} &  & {\scriptsize{}\citep{96}} & {\tiny{}\Checkmark{}} & {\scriptsize{}Measurement} & {\scriptsize{}A technique to map and analyze the Bitcoin broadcast
network}\tabularnewline
\cmidrule{3-6} \cmidrule{4-6} \cmidrule{5-6} \cmidrule{6-6} 
{\scriptsize{}modeling} &  & {\scriptsize{}\citep{98}} & {\tiny{}\Checkmark{}} & {\scriptsize{}Implementation} & {\scriptsize{}A model to discover Bitcoin network topology}\tabularnewline
\cmidrule{3-6} \cmidrule{4-6} \cmidrule{5-6} \cmidrule{6-6} 
 &  & {\scriptsize{}\citep{100}} & {\tiny{}\Checkmark{}} & {\scriptsize{}Implementation} & {\scriptsize{}A statistical model to analyze the connectivity of the
network}\tabularnewline
\cmidrule{3-6} \cmidrule{4-6} \cmidrule{5-6} \cmidrule{6-6} 
 &  & {\scriptsize{}\citep{101}} & {\tiny{}\Checkmark{}} & {\scriptsize{}Implementation} & {\scriptsize{}A topology discovery system to represent Bitcoin's network}\tabularnewline
\cmidrule{3-6} \cmidrule{4-6} \cmidrule{5-6} \cmidrule{6-6} 
 &  & {\scriptsize{}\citep{103}} & {\tiny{}\Checkmark{}} & {\scriptsize{}Measurement} & {\scriptsize{}A model to analyze network properties of BTC lightning
network}\tabularnewline
\cmidrule{3-6} \cmidrule{4-6} \cmidrule{5-6} \cmidrule{6-6} 
 &  & {\scriptsize{}\citep{105}} & {\tiny{}\Checkmark{}} & {\scriptsize{}Simulation} & {\scriptsize{}A probabilistic model to analyze Bitcoin protocol}\tabularnewline
\cmidrule{3-6} \cmidrule{4-6} \cmidrule{5-6} \cmidrule{6-6} 
 &  & {\scriptsize{}\citep{109}} & {\tiny{}\Checkmark{}} & {\scriptsize{}Simulation} & {\scriptsize{}Model delay and traffic overhead using Random Graph}\tabularnewline
\cmidrule{3-6} \cmidrule{4-6} \cmidrule{5-6} \cmidrule{6-6} 
 &  & {\scriptsize{}\citep{110}} & {\tiny{}\Checkmark{}} & {\scriptsize{}Measurement} & {\scriptsize{}Model behavior of network and dynamic topology}\tabularnewline
\cmidrule{2-6} \cmidrule{3-6} \cmidrule{4-6} \cmidrule{5-6} \cmidrule{6-6} 
 & {\tiny{}ETH} & {\scriptsize{}\citep{99}} & {\tiny{}\Checkmark{}} & {\scriptsize{}Measurement} & {\scriptsize{}A model to monitor and analyze Ethereum network}\tabularnewline
\cmidrule{3-6} \cmidrule{4-6} \cmidrule{5-6} \cmidrule{6-6} 
 &  & {\scriptsize{}\citep{102}} & {\tiny{}\Checkmark{}} & {\scriptsize{}Measurement} & {\scriptsize{}A model to analyze topological characteristics of ETH}\tabularnewline
\cmidrule{3-6} \cmidrule{4-6} \cmidrule{5-6} \cmidrule{6-6} 
 &  & {\scriptsize{}\citep{104}} & {\tiny{}\Checkmark{}} & {\scriptsize{}Implementation} & {\scriptsize{}A model to analyze and measure ETH P2P network}\tabularnewline
\cmidrule{3-6} \cmidrule{4-6} \cmidrule{5-6} \cmidrule{6-6} 
 &  & {\scriptsize{}\citep{106}} & {\tiny{}\Checkmark{}} & {\scriptsize{}Implementation} & {\scriptsize{}A model to study evolution and temporal characteristics
of ETH}\tabularnewline
\cmidrule{3-6} \cmidrule{4-6} \cmidrule{5-6} \cmidrule{6-6} 
 &  & {\scriptsize{}\citep{108}} & {\tiny{}\Checkmark{}} & {\scriptsize{}Measurement} & {\scriptsize{}A node discovery model for Ethereum's main network}\tabularnewline
\cmidrule{2-6} \cmidrule{3-6} \cmidrule{4-6} \cmidrule{5-6} \cmidrule{6-6} 
 & {\tiny{}BTC/ETH} & {\scriptsize{}\citep{97}} & {\tiny{}\Checkmark{}} & {\scriptsize{}Measurement} & {\scriptsize{}A model to measure the decentralization degree of blockchains}\tabularnewline
\cmidrule{2-6} \cmidrule{3-6} \cmidrule{4-6} \cmidrule{5-6} \cmidrule{6-6} 
 & {\tiny{}Algorand} & {\scriptsize{}\citep{111}} & {\tiny{}\Checkmark{}} & {\scriptsize{}Measurement} & {\scriptsize{}Analysis of network revolution and transaction activities}\tabularnewline
\cmidrule{2-6} \cmidrule{3-6} \cmidrule{4-6} \cmidrule{5-6} \cmidrule{6-6} 
 & {\tiny{}ETH2.0} & {\scriptsize{}\citep{107}} & {\tiny{}\Checkmark{}} & {\scriptsize{}Measurement} & {\scriptsize{}A model to analyze ETH 2.0 P2P network}\tabularnewline
\midrule 
{\scriptsize{}Workload} & {\tiny{}General} & {\scriptsize{}\citep{112}} & {\tiny{}\Checkmark{}} & {\scriptsize{}Simulation} & {\scriptsize{}Network characterization to predict its behavior}\tabularnewline
\cmidrule{3-6} \cmidrule{4-6} \cmidrule{5-6} \cmidrule{6-6} 
{\scriptsize{}characterization} &  & {\scriptsize{}\citep{113}} & {\tiny{}\Checkmark{}} & {\scriptsize{}Simulation} & {\scriptsize{}Network behavior prediction and performance evaluation}\tabularnewline
\cmidrule{2-6} \cmidrule{3-6} \cmidrule{4-6} \cmidrule{5-6} \cmidrule{6-6} 
 & {\tiny{}HLF} & {\scriptsize{}\citep{114}} & {\tiny{}\XSolid{}} & {\scriptsize{}Implementation} & {\scriptsize{}Workload characterization and evaluation of a BC implementation}\tabularnewline
\cmidrule{2-6} \cmidrule{3-6} \cmidrule{4-6} \cmidrule{5-6} \cmidrule{6-6} 
 & {\tiny{}BTC} & {\scriptsize{}\citep{126}} & {\tiny{}\Checkmark{}} & {\scriptsize{}Measurement} & {\scriptsize{}Characterization of Bitcoin transactions}\tabularnewline
\midrule 
{\scriptsize{}Benchmarking} & {\tiny{}ETH/HLF/Parity} & {\scriptsize{}\citep{125}} & {\tiny{}\XSolid{}} & {\scriptsize{}Implementation} & {\scriptsize{}Performance benchmarking using the workload of Blockbench}\tabularnewline
\cmidrule{2-6} \cmidrule{3-6} \cmidrule{4-6} \cmidrule{5-6} \cmidrule{6-6} 
{\scriptsize{}using} & {\tiny{}ETH/HLF} & {\scriptsize{}\citep{115}} & {\tiny{}\XSolid{}} & {\scriptsize{}Implementation} & {\scriptsize{}Performance analysis using varying numbers of transaction}\tabularnewline
\cmidrule{3-6} \cmidrule{4-6} \cmidrule{5-6} \cmidrule{6-6} 
{\scriptsize{}workloads} &  & {\scriptsize{}\citep{118}} & {\tiny{}\XSolid{}} & {\scriptsize{}Implementation} & {\scriptsize{}Performance benchmarking with BCTMark}\tabularnewline
\cmidrule{2-6} \cmidrule{3-6} \cmidrule{4-6} \cmidrule{5-6} \cmidrule{6-6} 
 & {\tiny{}Parity/Multichain} & {\scriptsize{}\citep{117}} & {\tiny{}\XSolid{}} & {\scriptsize{}Implementation} & {\scriptsize{}Performance evaluation using realistic BTC workload}\tabularnewline
\cmidrule{2-6} \cmidrule{3-6} \cmidrule{4-6} \cmidrule{5-6} \cmidrule{6-6} 
 & {\tiny{}HLF} & {\scriptsize{}\citep{116}} & {\tiny{}\XSolid{}} & {\scriptsize{}Implementation} & {\scriptsize{}Performance benchmarking using workload characterization}\tabularnewline
\midrule
{\scriptsize{}Traffic} & {\tiny{}General} & {\scriptsize{}\citep{123}} & {\tiny{}\Checkmark{}} & {\scriptsize{}Implementation} & {\scriptsize{}Model the network to predict the on-network load}\tabularnewline
\cmidrule{3-6} \cmidrule{4-6} \cmidrule{5-6} \cmidrule{6-6} 
{\scriptsize{}analysis} &  & {\scriptsize{}\citep{122}} & {\tiny{}\Checkmark{}} & {\scriptsize{}Simulation} & {\scriptsize{}Model traffic generated by synchronization protocols}\tabularnewline
\cmidrule{2-6} \cmidrule{3-6} \cmidrule{4-6} \cmidrule{5-6} \cmidrule{6-6} 
{\scriptsize{}and} & {\tiny{}BTC} & {\scriptsize{}\citep{119}} & {\tiny{}\Checkmark{}} & {\scriptsize{}Implementation} & {\scriptsize{}A model to discover propagation delay and forks}\tabularnewline
\cmidrule{3-6} \cmidrule{4-6} \cmidrule{5-6} \cmidrule{6-6} 
{\scriptsize{}modeling} &  & {\scriptsize{}\citep{120}} & {\tiny{}\Checkmark{}} & {\scriptsize{}Implementation} & {\scriptsize{}A model to predict the number of transactions of BTC}\tabularnewline
\cmidrule{3-6} \cmidrule{4-6} \cmidrule{5-6} \cmidrule{6-6} 
 &  & {\scriptsize{}\citep{121}} & {\tiny{}\Checkmark{}} & {\scriptsize{}Simulation} & {\scriptsize{}A model of the block arrival process}\tabularnewline
\cmidrule{3-6} \cmidrule{4-6} \cmidrule{5-6} \cmidrule{6-6} 
 &  & {\scriptsize{}\citep{124}} & {\tiny{}\XSolid{}} & {\scriptsize{}Implementation} & {\scriptsize{}Mathematically model traffic between Blockchain and
IoT devices}\tabularnewline
\bottomrule
\end{tabular*}
\raggedright{}{\scriptsize{}{*} Public permissionless}{\scriptsize\par}
\end{table}

\subsection{Benchmarking using workload}

In 2017, the authors in \citep{125} introduced Blockbench as an evaluation
framework designed to analyze performance of private blockchains.
It enables the integration of any private blockchain through straightforward
APIs and facilitates benchmarking against workloads derived from genuine
and artificial smart contracts. Subsequently, the authors utilized
Blockbench to perform an analysis of three blockchain systems---private
Ethereum, Parity, and Hyperledger---employing macro benchmarks and
micro benchmarks. Additionally, In 2020, Saingre et al. \citep{118}
introduced BCTMark, a benchmarking framework for blockchain systems
that incorporates features such as load generation and metrics collection.
To demonstrate its versatility, they conducted experiments on three
blockchain systems---Ethereum Clique, Ethereum Ethash and Hyperledger
Fabric. In 2017, Pongnumkul et al. \citep{115} discussed a performance
evaluation of Hyperledger Fabric and Ethereum, examining their capabilities
as two private blockchains under varying workloads. In 2018, Thakkar
et al. in \citep{116} characterized the transaction workload and
performance of Hyperledger Fabric to analyze performance parameters
and bottlenecks under various block sizes, policies, and network situations.
Using realistic workload to evaluate blockchains' performance, Oliveira
et al. in \citep{117} compared Parity and Multichain, two private
blockchains, focusing on transaction-related metrics using a workload
derived from the probability distribution of Bitcoin transaction arrival
times.

\subsection{Traffic analysis and modeling}

In \citep{119}, decker et al. examined Bitcoin's utilization of a
multi-hop broadcast to disseminate transactions and blocks across
the network for ledger updates. Subsequently, they presented a model
elucidating the presence of blockchain forks. Employing this model,
the authors confirmed the hypothesis that the primary reason for blockchain
forks is the propagation delay within the network.mThe goal of Bianconi
et al. in \citep{120} was to predict the forthcoming volume of transactions
within the Bitcoin network by leveraging its existing network structure.
This objective was similar to link prediction, but it specifically
aimed to estimate the expected quantity of connections (edges) within
the network over a defined time period. Modeling block intervals in
Bitcoin, Bowden et al. \citep{121} contradicted the initial claim
made in the Bitcoin paper proposing a homogeneous Poisson process
for blockchain arrivals. Their demonstration relied on empirical data
and stochastic analysis of block arrival patterns, revealing a different
reality. In 2018, Danzi et al. \citep{122} conducted a study focusing
on synchronization among IoT devices with the blockchain. They proposed
two protocols to synchronize and manage traffic between these networks,
analyzing the necessary bandwidth and time requirements for synchronization.
To examine the key features of blockchain networks traffic, Elagin
et al. \citep{123} introduced an analysis model that can predict
the network load, aiming to contribute towards ensuring secure and
high-quality data exchanges within this network in the future. Zhang
et al. \citep{124} introduced a traffic model for blockchain systems
within the context of IoT. They conducted an extensive analysis of
IoT input, blockchain network traffic, and self-similarity. They deployed
a lightweight blockchain similar to Bitcoin, utilizing Proof of Work
(PoW) connected to multiple IoT simulators to facilitate a range of
network traffic experiments.

\section{Challenges and future research directions}

In the preceding sections, we provide an overview of the performance
modeling landscape concerning public permissionless blockchains. Despite
significant advancements in the evaluation and modeling of blockchain
systems, several open questions persist. Using the information we've
gathered, we will discuss the challenges for performance modeling
of public permissionless blockchains in this section. We'll also suggest
some ideas for future research.

\subsection{Performance modeling and evaluation issues and future directions}

Performance modeling in public blockchains presents greater challenges
compared to private and permissioned blockchains due to the inherent
variability of network conditions and the decentralization of nodes.
While queueing models have been commonly used for performance modeling
in these networks, identifying the optimal modeling approach remains
an open question. This is primarily because queueing models encounter
limitations when applied to public blockchains, given their dynamic
and unpredictable nature. Overcoming these limitations necessitates
the advancement of more sophisticated modeling techniques capable
of capturing the unique attributes of public blockchain systems. 

Additionally, selecting the most suitable queue type for modeling
the performance of public blockchains poses another significant challenge.
For instance, in a study by Kasahara \citep{66}, the researcher investigated
the process of transaction confirmation in Bitcoin by employing three
types of queues and assessing their accuracy through simulation outcomes.

One of the challenges in evaluating and modeling the performance of
public blockchains is the measurement of certain performance parameters
like delays and transaction confirmation time. These parameters are
crucial for assessing blockchain performance, yet they pose difficulties
in measurement within public blockchains. For instance, various studies
have focused on analyzing transaction confirmation time using modeling
techniques. However, accurately determining this parameter's precise
value in a real-world public blockchain presents significant challenges.

As previously discussed, employing a modeling approach is considered
the most effective method for assessing the performance of public
blockchains. In this context, numerous researchers have conducted
studies on the performance modeling of Bitcoin and Ethereum, which
are comprehensively reviewed in this survey. However, the performance
modeling of recently developed public blockchains like Algorand and
Ethereum 2.0 has received minimal attention. Consequently, there exists
a promising avenue for future research to delve into the performance
modeling of these two blockchains, aiming to identify weaknesses and
propose adjustments accordingly.

Another requirement in this area is the absence of a dedicated simulator
designed for Ethereum 2.0 and Algorand. While one simulator has been
proposed by the authors in \citep{38}, Ethereum 2.0 has undergone
gradual changes since then, indicating an ongoing need for an updated
Ethereum 2.0 simulator. Consequently, the development of simulation
tools for newly introduced public blockchains emerges as a prospective
direction, offering support to other researchers engaged in performance
modeling of public blockchains.

\vspace{-.7cm} 

\subsection{Performance improvement issues and future directions}

As previously mentioned, transaction confirmation time stands out
as a critical performance metric in public blockchains. As illustrated
in Figure 2, this metric encompasses various components, such as transaction
propagation latency, transaction waiting latency, processing overheads,
block propagation latency, consensus latency, and finality delay,
all of which vary across different public blockchains. Modeling and
pinpointing these delays within public blockchains, to minimize each
latency individually, remains an unresolved issue in the realm of
performance modeling for public blockchains.

Another promising avenue for research involves analyzing the tradeoff
between performance and security in public blockchains, particularly
focusing on newly developed ones such as Algorand and Ethereum 2.0.
It is understood that, at some point, enhancing the performance of
public blockchains may require compromising the security of the system.
Investigating this balance and forecasting the implications on network
security when enhancing performance parameters represents a significant
research direction. Exploring this aspect of public blockchains can
potentially lead to the development of existing public blockchains
that maintain both acceptable performance and security simultaneously.
Furthermore, it would aid researchers in designing new public blockchains
with these desirable attributes.

\subsection{Traffic and workload issues and future directions}

During our review, we identified another promising avenue for research:
the characterization of transactions and traffic in newly developed
public blockchains. While certain papers have tackled similar challenges
within Bitcoin (e.g., \citep{126}), there remains a notable gap in
research about newly established public blockchains like Algorand
and Ethereum 2.0. Understanding and characterizing transactions and
incoming workloads within a public blockchain is of paramount importance
due to the substantial number of nodes and the inherent variability
of network conditions. Exploring this research direction could enable
researchers to create synthetic workloads for modeling and assessing
the performance of public blockchains in real-world scenarios. Consequently,
delving into the generation of synthetic traffic for public blockchains
emerges as another promising avenue in the realm of performance evaluation
for such systems.

\section{Conclusion}

After conducting an initial overview of existing surveys concerning
the performance evaluation of blockchains, we identified a crucial
gap. Specifically, there is an absence of a dedicated survey that
concentrates on evaluating the performance of blockchain systems,
specifically public permissionless blockchain systems. To address
this gap, we conducted an extensive survey concentrating on the most
advanced methods in performance modeling of blockchains, particularly
those models proposed for analyzing and evaluating the performance
of public permissionless blockchains. Within this survey, we presented
a taxonomy, ideas, and solutions for each category, aiming to provide
improved understanding and novel insights into utilizing modeling
approaches for predicting, analyzing, and evaluating the performance
of public permissionless blockchains. Additionally, recognizing the
significance of network topology, workload, and traffic characterization
in performance evaluation process, we delved into studies centered
on modeling blockchain network topology, analyzing traffic and workload
characteristics, and introducing blockchain benchmarking workloads
for the first time. We are of the opinion that our survey offers a
valuable guidance regarding the insights into evaluating blockchain
performance, catering to blockchain researchers, and a broader audience
of readers.

\bibliographystyle{2nd}
\nocite{*}
\bibliography{2nd}

\end{document}